\documentclass[preprint,authoryear, 13pt, a4paper]{elsarticle}
\usepackage{graphicx}
\usepackage{amssymb}
\usepackage{amsmath}
\usepackage{times,fancyhdr}
\usepackage{amsfonts}

\begin{document}
\begin{frontmatter}
\title{A survey and a molecular dynamics study on the (central) hydrophobic region of prion proteins}
\author{Jiapu Zhang$^\text{ab*}$, Feng Wang$^\text{a}$}
\address{$^\text{a}$Molecular Model Discovery Laboratory, Department of Chemistry \& Biotechnology,  Faculty of Science, Engineering \& Technology, Swinburne University of Technology,\\ 
Hawthorn Campus, Hawthorn, Victoria 3122, Australia;\\
$^\text{b}$Graduate School of Sciences, Information Technology and Engineering, \& Centre of Informatics and Applied Optimization, Faulty of Science, The Federation University of Australia,\\ 
Mount Helen Campus, Ballarat, Victoria 3353, Australia;\\
$^\text{*}$Email: jiapuzhang@swin.edu.au, j.zhang@federation.edu.au, jiapu\_zhang@hotmail.com\\
Tel: +61-3-9214 5596, +61-3-5327 6335, +61-423 487 360
}
\begin{abstract}
Prion diseases are invariably fatal neurodegenerative diseases that affect humans and animals. Unlike most other amyloid forming neurodegenerative diseases, these can be highly infectious. Prion diseases occur in a variety of species. They include the fatal human neurodegenerative diseases Creutzfeldt-Jakob Disease (CJD), Fatal Familial Insomnia (FFI), Gerstmann-Sträussler-Scheinker syndrome (GSS), Kuru, the bovine spongiform encephalopathy (BSE or `mad-cow' disease) in cattle, the chronic wasting disease (CWD) in deer and elk, and scrapie in sheep and goats, etc. Transmission across the species barrier to humans, especially in the case of BSE in Europe, CWD in North America, and variant CJDs (vCJDs) in young people of UK, is a major public health concern. Fortunately, scientists reported that the hydrophobic region of prion proteins (PrP) controls the formation of diseased prions. This article gives a detailed survey on PrP hydrophobic region and does molecular dynamics studies of human PrP(110--136) to confirm some findings from the survey.
\end{abstract}
\begin{keyword}
prion diseases; prion protein; central hydrophobic region; a detailed survey; molecular dynamics study
\end{keyword}
\end{frontmatter}

\section{Introduction}
\label{introduction}
``Mad cow disease" (i.e. Bovine Spongiform Encephalopathy (BSE)) belongs to a contagious type of Transmissible Spongiform Encephalopathies (TSEs). Scientists believe it is caused by Prions (the misfolding prion proteins) but they may have not yet solved the riddle of ``mad cow disease". This is due to a prion is neither a virus, a bacteria nor any microorganism so the disease cannot be caused by the vigilance of the organism immune system and it can freely spread from one species to another species. The humans exists the susceptibility of TSEs; for example, the human version of "mad cow disease" named Creutzfeldt-Jakob Disease (CJD) and variant CJD (vCJD) just happen randomly through infections of transplanted tissue or blood transfusions or consumption of infected beef products. Cat, mink, deer, elk, moose, sheep, goat, nyala, oryx, greater kudu, ostrich and many other species are also susceptible to TSEs. Scientists do not know the reason of TSEs.\\

Fortunately, scientists reported that the hydrophobic region 109--136 of prion proteins (PrP) controls the formation of diseased prions. This article firstly gives a detailed survey on PrP hydrophobic region; secondly, this article does molecular dynamics (MD) studies of human PrP(110--136) (PDB ID: 2LBG.pdb) to confirm some findings from the survey and to try to reveal something new.\\

It was reported that the hydrophobic region PrP(109--136) controls the formation of diseased prions: the normal PrP(113--120) AGAAAAGA palindrome is an inhibitor/blocker of prion diseases [\cite{brown_2000, holscher_etal1998}] and the highly conserved glycine-xxx-glycine motif PrP(119--131) can inhibit the formation of infectious prion proteins in cells [\cite{harrison_etal2010, cheng_etal2011, lee_etal2008}]. The first part of this article will review the research works on all the PrP hydrophobic regions.\\

The infectious diseased prion is thought to be an abnormally folded isoform (PrP$^{\text{Sc}}$) of a host protein known as the prion protein (PrP$^{\text{C}}$). The conversion of PrP$^{\text{C}}$ to PrP$^{\text{Sc}}$ occurs post-translationally and involves conformational change from a predominantly $\alpha$-helical protein to one rich in $\beta$-sheet amyloid fibrils. Much remains to be understood about how the normal cellular isoform of the prion protein PrP$^{\text{C}}$ undergoes structural changes to become the disease associated amyloid fibril form PrP$^{\text{Sc}}$. Such a structural conformational change may be amenable to study by MD techniques. The second part of this article will do MD studies on the hydrophobic region 110--136 of human PrP (HuPrP).\\

The rest of this paper is organized as follows. Firstly, we will review previous experimental research results in the laboratories (not on the computers) on the hydrophobic region of PrP listed in the PubMed of NCBI (http://www.ncbi.nlm.nih.gov/pubmed). Secondly, we will present MD simulation studies on HuPrP(110--136). Lastly, concluding remarks on PrP hydrophobic region are summarized.

\section{A detailed review on PrP hydrophobic region}
The highly and evolutionarily conserved PrP hydrophobic region has been considered essential to PrP conformational conversion. This section does a survey of the research work on {\sl prion hydrophobic region} listed in the PubMed database mainly from the molecular structure point of view.
\begin{itemize}
\item 2014: 
      \begin{itemize}
      \item Chu et al. (2014) equipped $\Delta$105-125, $\Delta$CR\_PrP (i.e. the PrP missing central hydrophobic region, a variant that is known to be highly neurotoxic in transgenic mice) with a C-terminal membrane anchor via a semisynthesis strategy and found the importance of the central hydrophobic domain in the membrane anchor in PrP-lipid interactions [\cite{chu_etal2014}].
      \item Daskalov et al. (2014) studied the HET-s prion (which owns a $\beta$-solenoid with a triangular hydrophobic core) and identified a region that modulates prion formation [\cite{daskalov_etal2014, saupe2011}]. Solid-state NMR data showed the hydrophobic core of HET-s(218--289) is rigid [\cite{lange_etal2009}]. 
      \item Xu et al. (2014) presented the structures of prion-like MAVS and found there are ``electrostatic interactions between neighboring strands and hydrophobic interactions within each strand" [\cite{xu_etal2014}].
      \item Baral et al. (2014) did promazine binding to mouse PrP and found the ``binding induces structural rearrangement of the unstructured region proximal to $\beta$1, through the formation of a `hydrophobic anchor'" and promazine "stabilizes the misfolding initiator-motifs such as the C terminus of $\alpha$2, the $\alpha$2-$\alpha$3 loop, as well as the polymorphic $\beta$2-$\alpha$2 loop" [\cite{baral_etal2014}].
      \item Coleman et al. (2014) studied the murine homologues (G113V and A116V, which lie in the hydrophobic domain of PrP) and concluded that ``the hydrophobic domain is an important determinant of PrP conversion" [\cite{coleman_etal2014, zhang2014a}].
      \item Mays et al. (2014) studied the C1 cleavage that occurs amino-terminal of PrP$^\text{C}$'s hydrophobic domain [\cite{mays_etal2014}].
      \end{itemize}      
\item 2013: 
      \begin{itemize}
      \item Wang et al. (2013) used [Au(bpy)Cl$_2$]PF$_6$ and [Au(dien)Cl]Cl$_2$ (where bpy is 2,2'-bipyridine and dien is diethylenetriamine), six prion peptides with either a His111-mutated or a Met109/112-mutated residue to investigate interaction and peptide aggregation [\cite{wang_etal2013}]. 
      \item Pimenta et al. (2013) described the NMR structure of ovine prion-like Doppel peptide (1--30) (PDB ID: 2M1J) and the interaction of this peptide with the conserved SRP54M hydrophobic groove [\cite{pimenta_etal2013}]. 
      \item The central domain (CD, 95–-133) of PrP$^{\text{C}}$ comprises the charge cluster (CC, 95-–110) and a hydrophobic region (HR, 112–-133); Vilches et al. (2013) reported that among CC, HR and CD only the CD peptide is neurotoxic and this peptide is able to activate caspase-3 and disrupt the cell membrane, leading to cell death [\cite{vilches_etal2013}]. 
      \item ``Among mutants spanning the region 95–-135, a construct lacking solely residues 105–-125 ($\triangle$105--125) had distinct properties when compared with the full-length prion protein 23–-231 or other deletions" [\cite{patel_etal2013}].
      \item Zweckstetter (2013) reported that ``prion stop mutants that accumulate in amyloidogenic plaque-forming aggregates fold into a $\beta$-helix" and residue 129 is located in the hydrophobic core of the $\beta$-helix, the trimer interface of a trimeric left-handed $\beta$-helix model is formed by residues L125, Y128 and L130 [\cite{zweckstetter2013}].
      \end{itemize}
\item 2012: 
      \begin{itemize}
       \item Lau et al. (2012) demonstrated that Shadoo (Sho) binds DNA and RNA in vitro via the arginine-rich (or alanine-rich) hydrophobic region of five tandem A/LAAG amino residue repeats R1-R5 (including tandem positively charged ``RGG boxes"), where is the most homologous region of Sho and PrP (but Sho sequences showed variability in the number of alanine residues) [\cite{lau_etal2012, daude_etal2010, stewart_etal2009}].
       \item Zhao et al. (2012) reported that ``two fixed missense mutations (102Ser$\rightarrow$Gly and 119Thr$\rightarrow$Ala), and three missense mutations (92Pro$>$Thr/Met, 122Thr$>$Ile and 139Arg$>$Trp) in the coding region presenting different (P$<$0.05) genotypic and allelic frequency distributions between cattle and buffalo" [\cite{zhao_etal2012}].
      \item 14--3--3 beta protein (highly abundant in brain, a biomarker for sCJD) interacts with the central hydrophobic amino acids 106–-126 of prion protein [\cite{jeong_etal2012}].
      \item The $\alpha$-cleavage within the central hydrophobic domain not only disrupts a region critical for both PrP toxicity and PrP$^\text{C}$ to PrP$^\text{Sc}$ conversion but also produces the N1 fragment that is neuroprotective and the C1 fragment that enhances the proapoptotic effect of staurosporine in one report and inhibits prion in another [\cite{liang_etal2012}].
      \item Zhang et al. (2012) found a new haplotype in a Sunite sheep and the sheep of Inner Mongolia in China have several haplotypes with the similar results of Stewart and Daude's: SPRN (shadowof prion protein homology) contained an alanine-rich sequence, which is homologous to a hydrophobic core with amyloidgenic characteristics in PrP [\cite{zhang_etal2012}].
      \item Sauve et al. (2012) published the NMR structure of HuPrP(110--136) in dodecylphosphocholine (DPC) micelles (PDB ID: 2LBG) [\cite{sauve_etal2012}].
      \end{itemize}
\item 2011:
      \begin{itemize}
      \item Biljan et al. (2011) published the NMR structure of HuPrP-M129 mutant V210I (85--231) (PDB ID: 2LEJ) including the unstructured N-terminal part (residues 90–-124) [\cite{biljan_etal2011}].
      \item Julien et al. (2011) reported the relative and regional stabilities of the hamster, mouse, rabbit, and bovine PrPs towards urea unfolding, and also investigated the effect of the S174N mutation in rabbit PrP$^\text{C}$ [\cite{julien_etal2011}].
      \item Shi et al. (2011) reported that point-mutations within the hydrophobic transmembrane region increase the amount of $^\text{Ctm}$PrP (a kind of PrP$^{Sc}$) in cells, such as human homologue A117V which is associated with GSS and G114V associated with gCJD, while the mutations outsides transmembrane region do not [\cite{shi_etal2011}].
      \end{itemize}
\item 2010:
      \begin{itemize}
      \item Wang et al. (2010) reported that ``the hydrophobic domain deletion mutant significantly weakened the hydrophobic rPrP-lipid interaction and abolished the lipid-induced C-terminal PK-resistance",  ``both disease-associated P105L and P102L mutations, localized between lysine residues in the positively charged region, significantly affected lipid-induced rPrP conversion" and ``the hydrophobic domain localized 129 polymorphism altered the strength of hydrophobic rPrP(recombinant mouse PrP)-lipid interaction" [\cite{wang_etal2010}].
      \item Ile et al. (2010) published the NMR structure of HuPrP-M129 mutant Q212P (PrP(90--231)) (PDB ID: 2KUN) and concluded that the Q212P mutation caused GSS syndrome might be due to the disruptions of the hydrophobic core consisting of $\beta$2–-$\alpha$2 loop and $\alpha$3 helix [\cite{ile_etal2010}].
      \item Biasini et al. (2010) revealed through the deleting residues 114–-121 ($\triangle$114--121) that the hydrophobic core region governs mutant prion protein aggregation and intracellular retention [\cite{biasini_etal2010}]. 
      \item Harrison et al. (2010) reported that there are similarities between A$\beta$ (Alzheimer's amyloid-$\beta$) and PrP in the segment of the three GxxxG repeats (where both A$\beta$ and PrP have the crucial residue Methionine located in the middle (GxMxG) of the last repeat) that controls prion formation, and found that minor alterations to this highly conserved region of PrP$^\text{C}$ drastically affect the ability of cells to uptake and replicate prion infection in both cell and animal bioassay [\cite{harrison_etal2010, zhang2014a}].
      \item Oliveira-Martins et al. (2010) reported that ``PrP$^\text{C}$ undergoes extensive proteolysis at the $\alpha$ site (109K$\downarrow$H110)" and ``C1 prevalence was unaffected by variations in charge (PrP(90--110)) and hydrophobicity (PrP(110--130)) of the region neighbouring the $\alpha$-cleavage site, and by substitutions of the residues in the palindrome that flanks this site, instead, $\alpha$-cleavage was size-dependently impaired by deletions within the domain PrP(106–-119)" [\cite{oliveira-Martins_etal2010}].
      \item The hydrophobic core (HC), a transmembrane domain, harbors the most highly conserved regions of PrP$^\text{C}$. A mutation in HC is associated with prion disease resulting in an enhanced formation of a transmembrane form ($^\text{Ctm}$PrP). Lutz et al. (2010) created a set of mutants carrying microdeletions of 2–8 amino acids within HC between position 114 and 121 ($\triangle$114--121), and showed that these mutations display reduced propensity for transmembrane topology and HC might function as recognition site for the protease(s) responsible for PrP$^\text{C}$ $\alpha$-cleavage[\cite{lutz_etal2010}]. They also found that the mutant G113V displayed increased $^\text{Ctm}$PrP topology and decreased $\alpha$-cleavage in their in vitro assay and concluded that HC represents an essential determinant for transmembrane PrP topology, whereas the high evolutionary conservation of this region is rather based upon preservation of PrP$^\text{C}$ $\alpha$-cleavage, thus highlighting the biological importance of this cleavage.
      \item Valensin et al. (2010) reported that HuPrP possesses two copper binding sites localized at His-96 and His-111 in the so called ``amylodogenic" or neurotoxic region (residues 91-–126) and chicken PrP possesses a similar region (PrP(105--140)) containing two His (His-110 and His-124) and an identical hydrophobic tail of 15 amino acids rich in Ala and Gly [\cite{valensin_etal2010}].
      \item Turi et al. (2010) studied HuPrP(91--115), HuPrP(76--114)H85A and HuPrP(84--114)H96A and found that His96 predominates almost completely for nickel(II) ions, while His85 and His111 predominate copper(II) ions [\cite{turi_etal2010, joszai_etal2012, cui_etal2014}].
      \end{itemize} 
\item 2009:
      \begin{itemize}
      \item Julien et al. (2009) reported that ``strong hydrophobic interactions between helices $\alpha$1 and $\alpha$3, and between $\alpha$2 and $\alpha$3, stabilize these regions even at very high concentrations of urea" but this result is for PrP structural region 121--230 of bovine [\cite{julien_etal2009}].
      \item Fei et al. (2009) reported Preceding with a hydrophobic residue cysteine, instead of a charged residue, ``caused the rate of assembly into fibrils to increase greatly for both peptides and full-length Ure2p", and concluded that ``disulfide bond formation significantly accelerates the assembly of Ure2p fibrils because of the proximity of a potential amyloid stretch" [\cite{fei_etal2009}].
      \item Tseng et al. (2009) reported that hydrophobic does not play a major role in the unfolding of $\alpha$-helix 1 of PrP$^\text{C}$ [\cite{tseng_etal2009}] but In the mouse PrP (PDB ID: 1AG2) there are nine hydrophobic residues and five non-polar Glycines in $\beta$1-strand: GLGGYMLGSAMSRPMIHFGN(PrP(124--142)), five hydrophobic residues in $\beta$2-strand: N153MYRYPNQV YYRPVD167 (PrP(153--167)).
      \end{itemize}
\item 2008:
      \begin{itemize}
      \item Margittai et al.(2008) found that, to maximize the hydrophobic contact surface, most fibrils (that contain an extensive core region of about 20 amino acids or more) share a common parallel in-register arrangement of $\beta$-strands [\cite{margittai_etal2008}].
      \item Ciccotosto et al. (2008) reported ``the major region of neurotoxicity has been localized to the hydrophobic domain located in the middle of the PrP sequence" [\cite{ciccotosto_etal2008}].
      \end{itemize}
\item 2007:
      \begin{itemize}
      \item Owen et al. (2007) reported ``the protease thermolysin cleaves at the hydrophobic residues Leu, Ile, Phe, Val, Ala, and Met, residues that are absent from the protease accessible aminoterminal region of PrP$^\text{Sc}$" [\cite{owen_etal2007}].
      \item Ott et al. (2007) reported ``the presence of small hydrophobic amino acids such as Val and Ile was insufficient to promote re-orientation. Only Met and Leu, large hydrophobic amino acids, could promote transmembrane domain inversion" [\cite{ott_etal2007}]. 
      \item Sakudo et al. (2007) reported that neurotoxic signals of aged PrP(106--126) are mediated by N-terminal half of the hydrophobic region (HR) but not the octapeptide repeat (OR) of PrP [\cite{sakudo_etal2007}]. 
      \item Berti et al. (2007) reported the copper(II) binding result of the 91--120 region of HuPrP: ``room-temperature NMR spectroscopy data were consistent with the binding site centered on His111 being approximately fourfold stronger than that centered on His96, low-temperature EPR spectroscopy results yielded evidence for the opposite trend", and the MD results showed that ``Met112 approaching the metal at room temperature, a process that is expected to stabilize the His111-centered binding site through hydrophobic shielding of the metal coordination sphere" [\cite{berti_etal2007}].
      \item Harrison et al. (2007) summarized the GxxxG PrPctm motif GAVVGGLGGYMLG and made a mutation G131V leading to the possibility that this motif may be relevant to the actions of TM-PrP [\cite{harrison_etal2007}].
      \end{itemize}
\item 2006:
      \begin{itemize}
      \item Lee et al. (2006) found that ``the OR (octapeptide repeat region, PrP(51--90)) and N-terminal half of HR (hydrophobic region, PrP(112--145)) of PrP retains anti-apoptotic activity similar to full-length PrP" [\cite{lee_etal2006}].  
      \end{itemize}
\item 2005:
      \begin{itemize}
      \item Sakudo et al. (2005a) reported that ``OR and N-terminal half of the HR were required for the inhibitory effect of PrP(113-–132) but not STI1 (stress-inducible protein 1) pep.1" and ``mediate the  the action of STI1 upon cell survival and upregulation of SOD (superoxide dismutase) activity" [\cite{sakudo_etal2005a}]. In 2005, Sakudo et al. (2005b) also reported that ``Removal of the OR (PrP(53--94)) enhances apoptosis and decreases SOD activity. Deletion of the N-terminal half of HR (PrP(95--132)) abolishes its ability to activate SOD and to prevent apoptosis, whereas that of the C-terminal half of HR (PrP(124–-146)) has little if any effect on the anti-apoptotic activity and SOD activation" [\cite{sakudo_etal2005b}]. 
      \item Gaggelli et al. (2005) did NMR and EPR studies on the interaction of the HuPrP(106--126) (KTNMKHMAGAAAAGAVVGGLG) with copper(II), manganese(II) and zinc(II) at pH 5.7 and concluded that ``the hydrophobic C-terminal region was not affecting the copper-binding (of His111) properties of the peptide and this hydrophobic tail is left free to interact with other target molecules" [\cite{gaggelli_etal2005}].
      \end{itemize} 
\item 2004:
      \begin{itemize}
      \item  Kuznetsov et al. (2004) reported that ``PrP(114–-125) and the C terminus of helix B may be considered as primary candidates for sites involved in conformational transition from PrP$^\text{C}$ to PrP$^\text{Sc}$" and ``most PrP mutations associated with neurodegenerative disorders increase local hydrophobicity" [\cite{kuznetsov_etal2004}].
      \item Ott et al. (2004) reported that ``the N terminal and hydrophobic regions of the signal sequence affect (membrane) integration (of priton protein) most significantly. Mutations in either region result in a considerable increase in the number of chains that integrate" [\cite{ott_etal2004}].
      \item  Haire et al. (2004) reported that the strand of residues 129–-131 in sheep PrP crystal structure is involved in lattice contacts about a crystal dyad to generate a four-stranded intermolecular $\beta$-sheet between neighbouring molecules [\cite{haire_etal2004}].  
      \end{itemize}
\item 2003:
      \begin{itemize}
      \item Susceptibility to scrapie is primarily controlled by polymorphisms in the ovine PrP gene (PRNP) and Seabury et al. (2003) reported a polymorphism of ovine PrP at Proline 116, flanking the cleavage site between Lys112 and His113, and A117V (causing Gerstmann-Strausler-Scheinker syndrome) [\cite{seabury_etal2003}].
      \item Saez-Cirion et al. (2003) reported that the hydrophobic internal region 130GAVVGGLGGYMLGSAMSR147 of bovine PrP shares structural and functional properties with HIV type 1 fusion peptide [\cite{saez-cirion_etal2003}]. The 121KHVAGAAAAGAVVGGLGGYMLGSAMSR147 transmembrane region (BovinePrP(tm)) has mainly a helical structure but also containing some random coil (upon addition of calcium, the random coils disappear while the helical conformation remains) [\cite{saez-cirion_etal2003}]. Amyloid-type fibers PrP$^\text{Sc}$ will be formed, if in the absence of membrane prion sequence [\cite{saez-cirion_etal2003}].
      \item Tcherkasskaya et al. (2003) reported that hydrophobic interactions between side chains of the peptide variants PrP(104--122) (containing a polar head KTNMKN followed by a long hydrophobic tail MAGAAAAAGAVV) and PrP(109--122) prevent the formation of the rigid $\beta$-sheet structures [\cite{tcherkasskaya_etal2003}]. 
      \item Premzl et al. (2003) reported that the alignment of the hydrophobic segment shows strong conservation across all PrPs and Shadoos; 12 of the 20 residues of the hydrophobic region are identical or almost identical, and another 6 are conserved hydrophobic [\cite{premzl_etal2003}].
      \item Cui et al. (2003) investigated the deletions of the hydrophobic domain (PrP$\Delta$(112-–119), PrP(112-–136), PrP(135-–150)) and found that the conserved hydrophobic core region is a critical domain for the activity of PrP [\cite{cui_etal2003}].
      \item Kourie et al. (2003) reported that copper modulates the ion channels of PrP[106--126] mutant prion peptide fragments, the hydrophobic core AGAAAAGA is not a ligand Cu(2+) site but the Cu(2+)-binding site is located at M(109) and H(111) of prion fragment PrP[106--126] [\cite{kourie_etal2003}].
      \item In mouse PrP, the selected rabbit-specific mutations Asn99Gly, Leu108Met, Asn173Ser, and Val214Ile significantly interfere with the conversion of PrP$^\text{C}$ to PrP$^\text{Sc}$ [\cite{vorberg_etal2003}].
      \end{itemize}
\item 2002:
      \begin{itemize}
      \item Suzuki et al. (2002) reported that Fugu PrP-like has the PrP-conserved hydrophobic region of the Xenopus PrP (but doppel gene lacks PrP conserved hydrophobic region) [\cite{suzuki_etal2002}].
      \item The amino acid sequence AGAAAAGA, comprising residues 112-–119 of the murine PrP$^\text{C}$, has been shown to be amyloidogenic and evolutionarily conserved [\cite{zhang2014b}]. To assess the effect of mutations at and around this hydrophobic sequence on protease resistance, Wegner et al. (2002) replaced the sequence either by alanines or by glycines and, in a third mutant, a large part surrounding this region was removed. At last Wegner et al. (2002) concluded that ``mutations in the central hydrophobic region lead to immediate alterations in PrP structure and processing" [\cite{wegner_etal2002}].
      \item Chen et al. (2002) reported ``one O-linked sugar at Ser135 can affect the coil-to-beta structural transition of the prion peptide" but at Ser132 the effect is opposite [\cite{chen_etal2002}], where PrP$^\text{C}$(132-–140) domain is closely related to the prion propagation and 132--140 portion of PrP$^\text{C}$ is a logical target for development of anti-prion drugs [\cite{peretz_etal2001}].
      \end{itemize}  
\item 2001: 
      \begin{itemize}
      \item Dragani et al. (2001) showed that, in the hydrophobic core of all GST (glutathione S-transferase) and related protein (including the yeast prion protein Ure2), at the beginning of $\alpha$6-helix of GST domain II, a N-capping box (S/TxxD) and an hydrophobic staple motif are strictly conserved in all GSTs and GST-related proteins [\cite{dragani_etal2001, khan_etal2010, sweeting_etal2013}].
      \item Holscher et al. (2001) reported that ``deletion of each of the two hydrophobic regions in PrP (i.e. $\Delta$PrP(112--121), $\Delta$PrP(231-254)) revealed that the C-terminally located hydrophobic region (transmembrane2, PrP(213--254)) can function as second signal sequence in PrP" [\cite{holscher_etal2001}].
      \item A direct correlation between the PrP amino acid sequence and TSE incubation time was demonstrated in transgenic mice expressing mouse PrP with amino acid substitutions Val111Met, Leu108Met, Pro101Leu [\cite{supattapone_etal2001, barron_etal2001, barron_etal2003}].
      \item Knaus et al. (2001) observed the N-terminal residues PrP(119–-124) in the electron density of the crystal structure of HuPrP dimer (but the PrP(119--124) region is disordered in the NMR structure of PrP monomers) [\cite{knaus_etal2001}]. 
      \item Laws et al. (2001) reported the solid-state NMR structure of MoPrP(89–-143, P101L) [\cite{laws_etal2001}].
      \end{itemize} 
\item 1999:
      \begin{itemize}
      \item Ragg et al. (1999) obtained the two-dimensional $^1$H NMR structure of PrP(106--126) under the following solvent conditions: deionized water/2,2,2-trifluoroethanol 50:50 (v/v) and dimethylsulfoxide; and the data were analyzed by restrained MD calculations [\cite{ragg_etal1999}]. ``In deionized water at pH 3.5, the peptide adopted a helical conformation in the hydrophobic region spanning residues Met112--Leu125, with the most populated helical region corresponding to the Ala115--Ala119 segment ($\thickapprox$ 10\%). In trifluoroethanol/H$_2$O, the $\alpha$-helix increased in population especially in the Gly119--Val122 tract ($\thickapprox$ 25\%). The conformation of this region was found to be remarkably sensitive to pH, as the Ala120--Gly124 tract shifted to an extended conformation at pH 7. In dimethylsulfoxide, the hydrophobic cluster adopted a prevalently extended conformation. For all tested solvents the region spanning residues Asn108--Met112 was present in a `turn-like' conformation and included His111, situated just before the starting point of the $\alpha$-helix. Rather than by conformational changes, the effect of His111 is exerted by changes in its hydrophobicity, triggering aggregation. The amphiphilic properties and the pH-dependent ionizable side-chain of His111 may thus be important for the
modulation of the conformational mobility and heterogeneity of PrP(106--126)" [\cite{ragg_etal1999}].
      \item Kanyo et al. (1999) determined the X-ray crystallographic atomic-resolution structures of Syrian hamster PrP(104--113) binding Fab 3F4 [\cite{kanyo_etal1999}]. ``The peptide binds in a U-shaped groove on the Fab surface, with the two specificity determinants, Met109 and Met112, penetrating deeply into separate hydrophobic cavities formed by the heavy and light chain complementarity-determining regions, and two intrapeptide hydrogen bonds are observed" [\cite{kanyo_etal1999}].
      \item Liu et al. (1999) reported some NMR information for the Syrian hamster PrP(90--123): ``the N-terminus (residues 90--119) is largely unstructured despite some sparse and weak medium-range NOEs implying the existence of bends or turns. The transition region between the core domain and flexible N-terminus, i.e., residues 113--128, consists of hydrophobic residues or glycines and does not adopt any regular secondary structure in aqueous solution. There are about 30 medium- and long-range NOEs within this hydrophobic cluster, so it clearly manifests structure" in 1B10.pdb [\cite{liu_etal1999}].
      \end{itemize} 
\item 1997:
      \begin{itemize}
      \item Smith et al. (1997) compared CD (circular dichroism) data from two peptides corresponding to the hydrophobic region PrP(106--136) which contained either methionine or valine at position 129 and concluded that ``there was no detectable difference between the CD spectra of these peptides irrespective of the solvent conditions we used" [\cite{smith_etal1997}]. 
      \end{itemize}
\item 1996:
      \begin{itemize}
      \item Inouye et al. (1996) revealed the hydrophobic core in prion by X-ray diffraction as the beta-silk structure [\cite{inouye_etal1996}].  
      \end{itemize}
      \item 1995:
      \begin{itemize}
      \item  Priola et al. (1995) found that ``a single hamster PrP amino acid Met138 blocks conversion to protease-resistant PrP in scrapie-infected mouse neuroblastoma cells" (while a mouse specific isoleucine Ile138 facilitated the conversion of mouse PrP$^\text{C}$ into PrP$^\text{Sc}$) [\cite{priola_etal1995}]. They also found homology at PrP(112--138) was required for mouse PrP$^\text{C} \rightarrow$ PrP$^\text{Sc}$ conversion [\cite{priola_etal1995, moore_etal2005}].
      \end{itemize}    
\item 1994:
      \begin{itemize}
      \item  De Fea et al. (1994) reported that ``a charged, extracytoplasmic region, termed the Stop Transfer Effector (STE) sequence, has been shown to direct then ascent translocating chain to stop at the adjoining hydrophobic domain to generate the first membrane-spanning hydrophobic region PrP(112--143)" [\cite{defea_etal1994}].
      \end{itemize}
\item 1993:
      \begin{itemize}
      \item Harris et al. (1993) proposed that the cleavage is lying within a region of 24 amino acids that is identical in chicken PrP and mammalian PrP and is representing a major processing event that may have physiological as well as pathological significance [\cite{harris_etal1993}].
      \end{itemize}        
\item 1992:
      \begin{itemize}
      \item Muramoto et al. (1992) found that $\Delta$PrP(95--107), $\Delta$PrP(108--121), and $\Delta$PrP(122--140) inhibit the PrP$^\text{C}$ $\rightarrow$PrP$^\text{Sc}$ conversion [\cite{muramoto_etal1992}]. Kuwata et al. (2003) furthermore reduced the region of residues and found that mouse PrP(106--126) can form amyloid-like fibrils (``the fibrils contain $\approx$ 50\% $\beta$-sheet structure, and strong amide exchange protection is limited to the central portion of the peptide spanning the palindromic sequence VAGAAAAGAV") [\cite{kuwata_etal2003}]. PrP(90--140) might be the region of interactions between PrP$^\text{C}$ and PrP$^\text{Sc}$ [\cite{viles_etal2001}].
      \end{itemize}
\item 1987:
      \begin{itemize} 
      \item Liao et al. (1987) cloned and sequenced rat prion-related protein (PrP) cDNA, and showed a 23 amino acids hydrophilic region that extends to amino acid position 122, and the numerous G-G-G-X repeats, where X is a hydrophobic amino acid, may function in $\beta$-pleated sheet amyloid formation [\cite{liao_etal1987}].
      \item Bazan et al. (1987) analyzed the primary PrP sequence hydrophobicities to detect potential amphipathic regions and several hydrophobic segments, a proline- and glycine-rich repeat region and putative glycosylation sites are incorporated into a PrP model [\cite{bazan_etal1987}].
       \end{itemize}     
\end{itemize}  
 
\noindent Throughout the above review on the recent research advances of PrP hydrophobic region, we noticed that hydrophobic region PrP(109-136) controls the formation of diseased prions PrP$^\text{Sc}$ conformationally  changed from  PrP$^\text{C}$. Thus, it is necessary to do MD studies of the PrP(110--136) hydrophobic region in the next section.

\section{MD (molecular dynamics) studies on PrP hydrophobic region 110--136}
The NMR structure of HuPrP(110-136) in dodecylphosphocholine micelles was known (2LBG.pdb). We did the MD simulations for this PDB template in [\cite{zhang_etal2013}] but we have not published any figure of our MD results. This section will illuminate our MD results and their discussions with numerous figures.\\ 

The 5 ns' MD simulations passed (1) successful optimizations of the NMR structure and its 17 mutants (G114V, A117V, G119A, G119L, G119P, A120P, G123A, G123P, G124A, L125A, G126A, G127A, G127L, M129V, G131A, G131L, G131P), (2) the longer enough equilibration phase: the PRESS, VOLUME (DENSITY) and RMSD (Figs. 1--18) were sufficiently stable, and (3) the successful production phase (Figs. 1--18). In Figs. 1--18, H is the $\alpha$-helix, B is the residue in isolated $\beta$-bridge, E is the extended strand, G is the 3-helix or 3$_\text{10}$ helix, I is the $\pi$-helix, T is the hydrogen bonded turn, S is the bend, and the dashed lines denote hydrogen bonds (HBs).\\

We found three main HBs GLY/ALA/LEU/PRO131--SER135--ARG136, TYR128--SER132 have high occupied rates in the N-terminal of PrP structural region; this shows that PrP(110--124) region of PrP(110--136) (2LBG.pdb) is very unstable. We also found that there is a salt bridge (SB) between HIS111--LYS110 with the occupied rate 100\% for 13 models. Specially for the mutants G127L, M129V, G131A, and G131L, there is another SB between HIS111--136ARG linking the head and the tail of HuPrP(110--136). Seeing the snapshots of 3 ns, 4 ns and 5 ns of mutant G127L in Fig. 13, 4 ns and 5 ns of mutant M129V in Fig. 15, 3 ns, 4 ns and 5 ns of mutant G131A in Fig. 16, and 3 ns, 4 ns and 5 ns of mutant G131L in Fig. 17, we may know the SB HIS111--ARG136 makes these snapshots looking like a ``hairpin". The residue HIS111 is an important residue in PrP(110--136). Along with the unfolding of $\alpha$-helical structure of all these 13 models, we found many hydrophobic packings (HPs) disappeared except for some fundamental ones MET134--ALA133, LEU130--129MET, VAL122--VAL121--ALA120, ALA118--ALA117--ALA116--ALA115, and ALA113--MET112 with the occupied rate 100\%, where ALA118--ALA117--ALA116--ALA115 are in the core of the palindrome AGAAAAGA and this might imply to us the hydrophobic core is very hard to break and this palindrome really has enormous potential to be amyloid fibrils.

\section{Conclusion}
It was reported that PrP hydrophobic region controls the formation into diseased prions. This article firstly reviewed on the recent research advancements of PrP hydrophobic region. Secondly, this article gave explanations on the PrP hydrophobic region in view of protein three-dimensional (3D) structures and their structural dynamics by numerous figures that illuminated the variations of protein 3D structures, protein secondary structures, and the snapshots at 0 ns, 1 ns, 2 ns, 3 ns, 4 ns and 5 ns of the MD simulations. The structural bioinformatics presented in this article can be used as a reference in 3D images for laboratory experimental works to study PrP hydrophobic region.

\section*{Acknowledgments}
{\small This research has been supported by a Victorian Life Sciences Computation Initiative (VLSCI) grant numbered VR0063 on its Peak Computing Facility at the University of Melbourne, an initiative of the Victorian Government of Australia.}

\bibliographystyle{elsarticle-harv}


\begin{figure}[h!] \label{Fig02}
\centerline{
\includegraphics[width=6.4in]{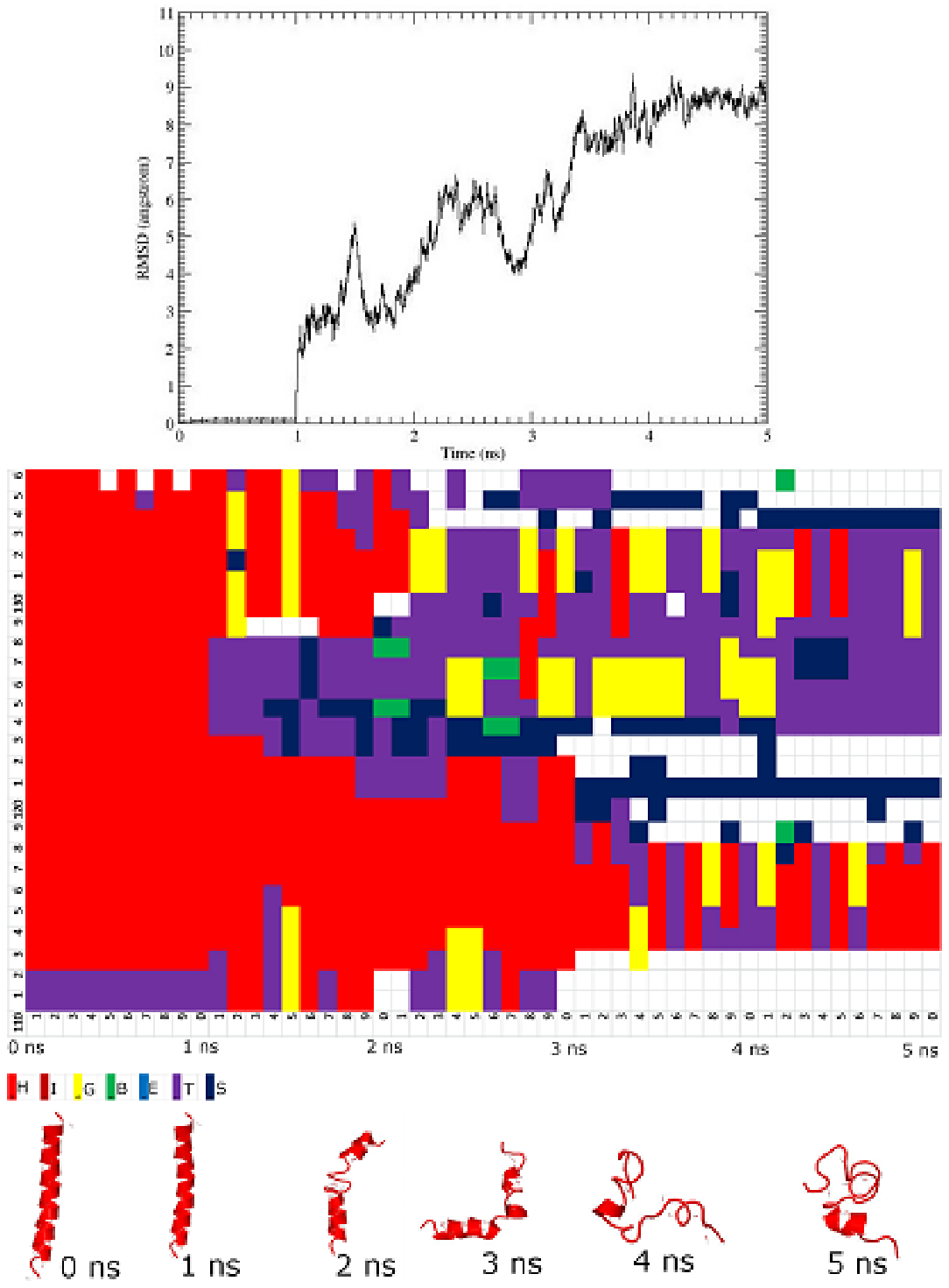}
}
\caption{{\bf HuPrP(110-136)} Variations of (A) the root-mean-squared deviations (RMSD), (B) the secondary structures, and (C) the respective snapshots at 0 ns, 1 ns, 2 ns, 3 ns, 4 ns and 5 ns of the MD simulations for HuPrP(110-136).}
\end{figure}

\begin{figure}[h!] \label{Fig03}
\centerline{
\includegraphics[width=6.4in]{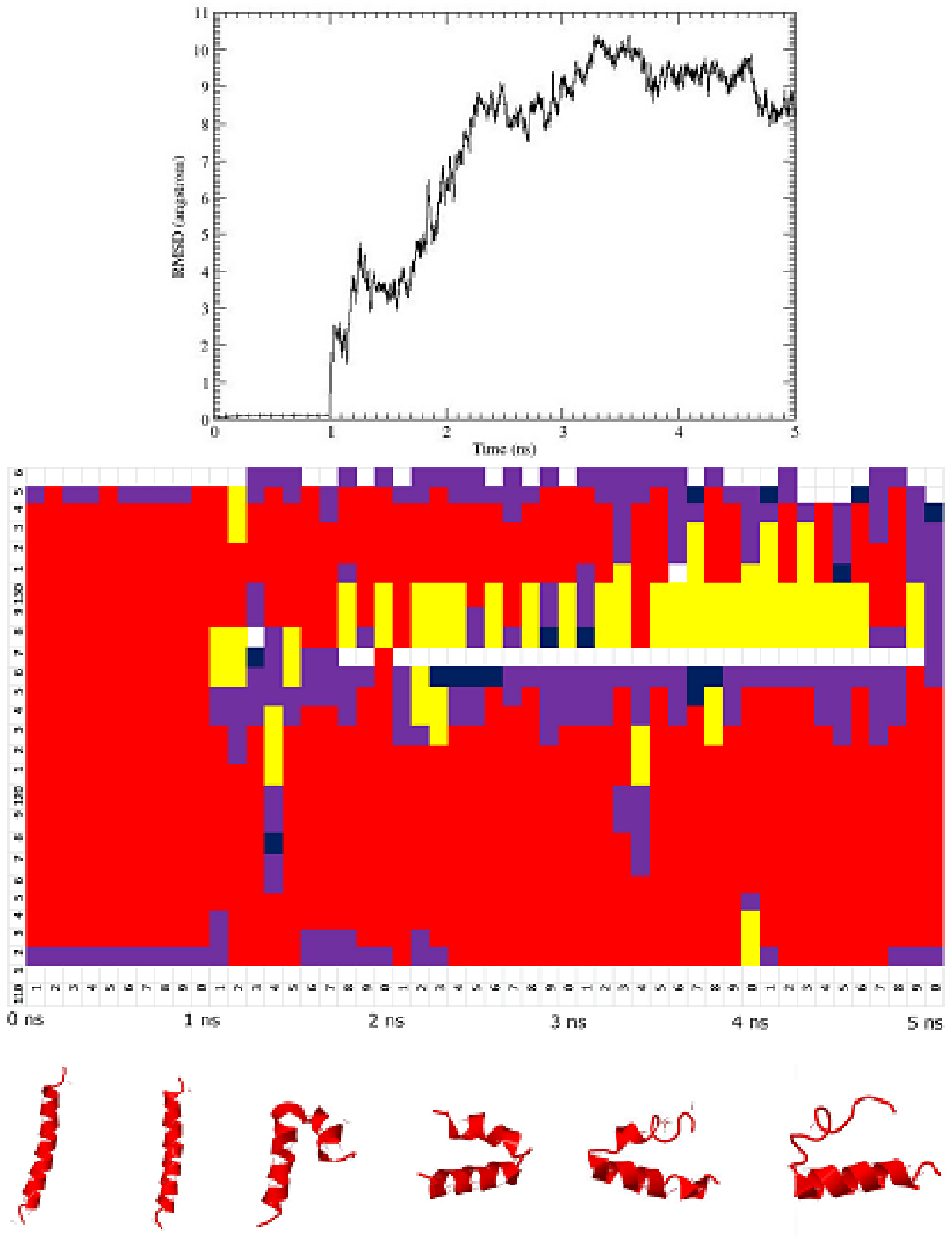}
}
\caption{{\bf Mutant G114V} Variations of (A) the root-mean-squared deviations (RMSD), (B) the secondary structures, and (C) the respective snapshots at 0 ns, 1 ns, 2 ns, 3 ns, 4 ns and 5 ns of the MD simulations for mutant G114V.}
\end{figure}

\begin{figure}[h!] \label{Fig04}
\centerline{
\includegraphics[width=6.4in]{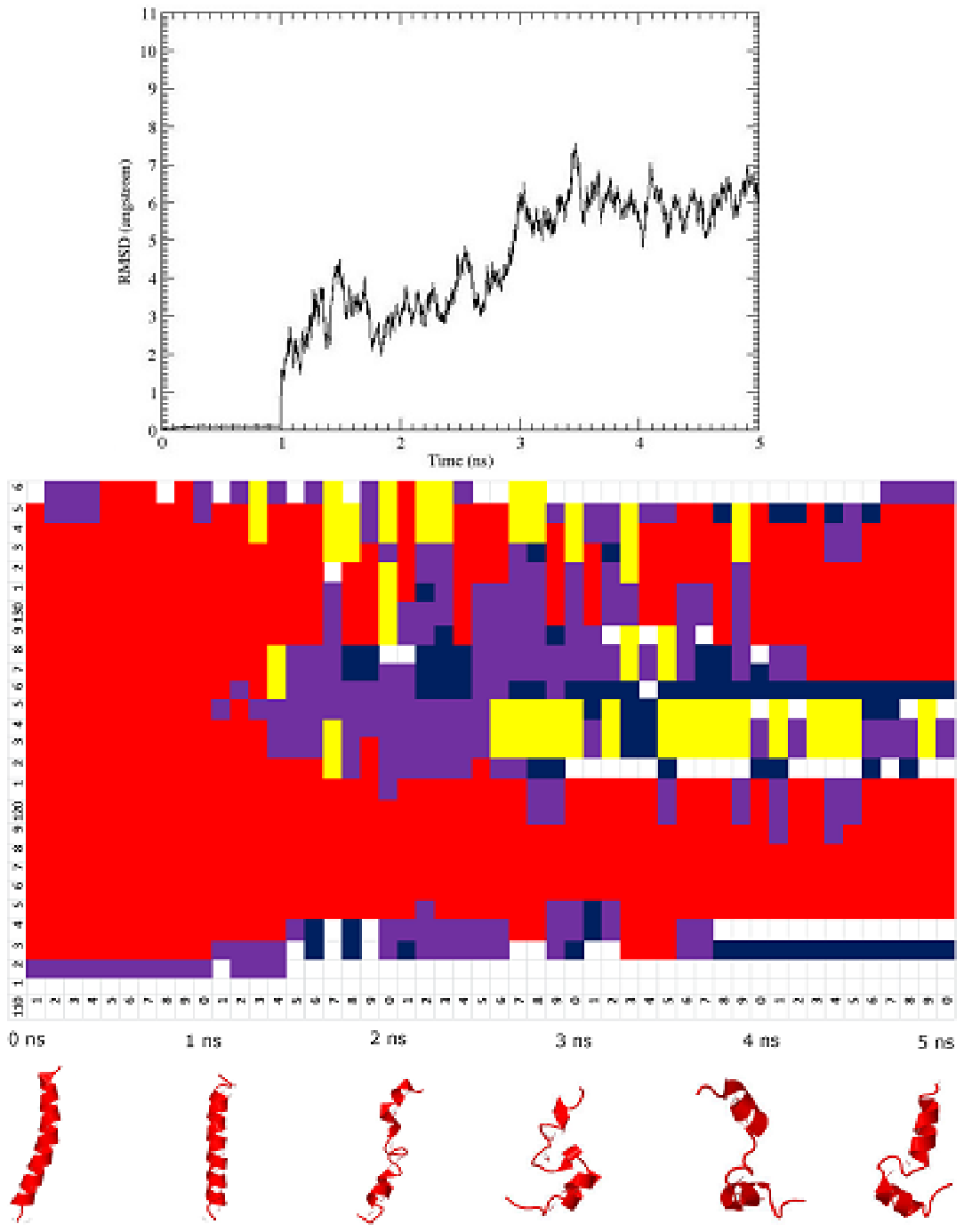}
}
\caption{{\bf Mutant A117V} Variations of (A) the root-mean-squared deviations (RMSD), (B) the secondary structures, and (C) the respective snapshots at 0 ns, 1 ns, 2 ns, 3 ns, 4 ns and 5 ns of the MD simulations for mutant A117V.}
\end{figure}

\begin{figure}[h!] \label{Fig05}
\centerline{
\includegraphics[width=6.4in]{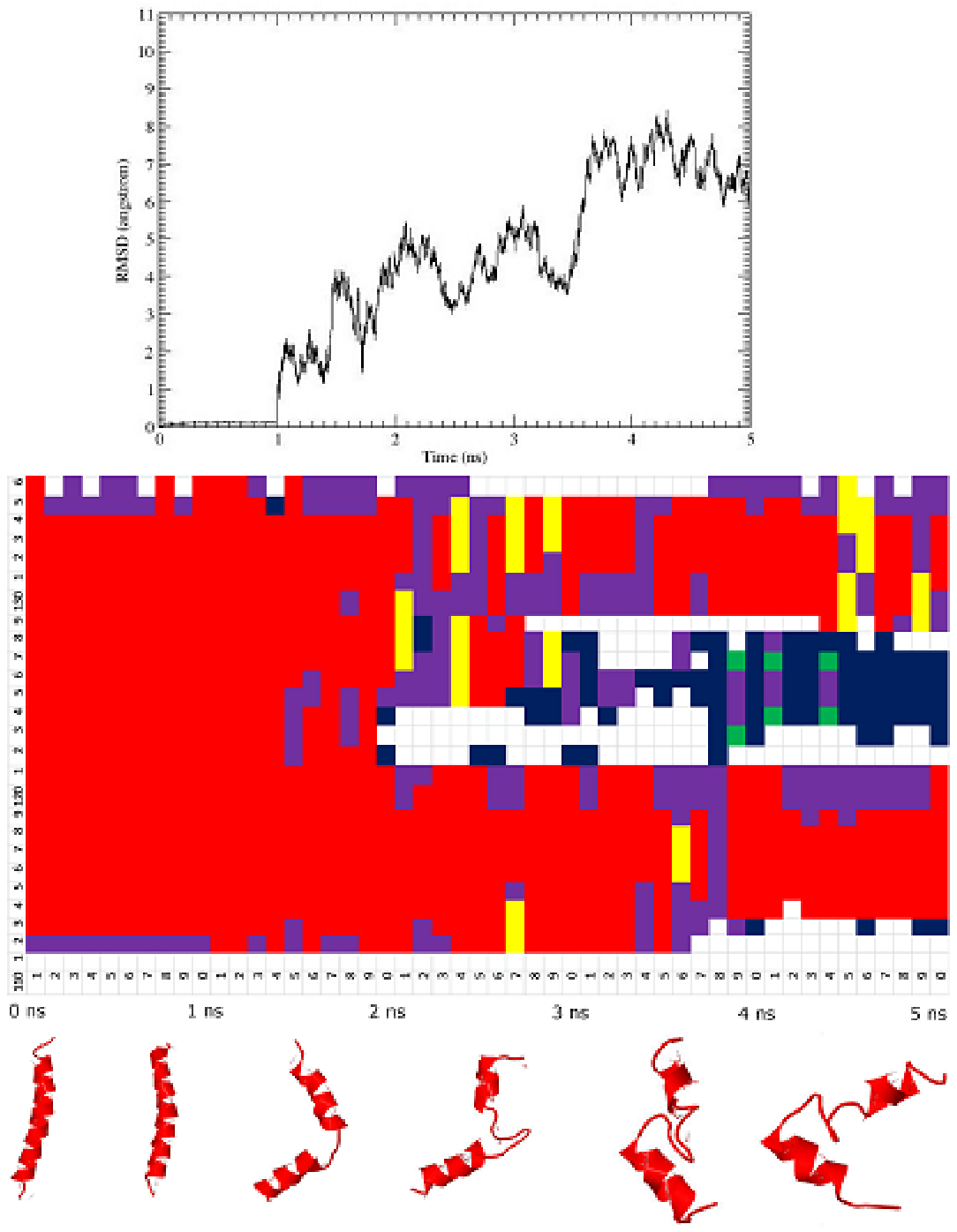}
}
\caption{{\bf Mutant G119A} Variations of (A) the root-mean-squared deviations (RMSD), (B) the secondary structures, and (C) the respective snapshots at 0 ns, 1 ns, 2 ns, 3 ns, 4 ns and 5 ns of the MD simulations for mutant G119A.}
\end{figure}

\begin{figure}[h!] \label{Fig06}
\centerline{
\includegraphics[width=6.4in]{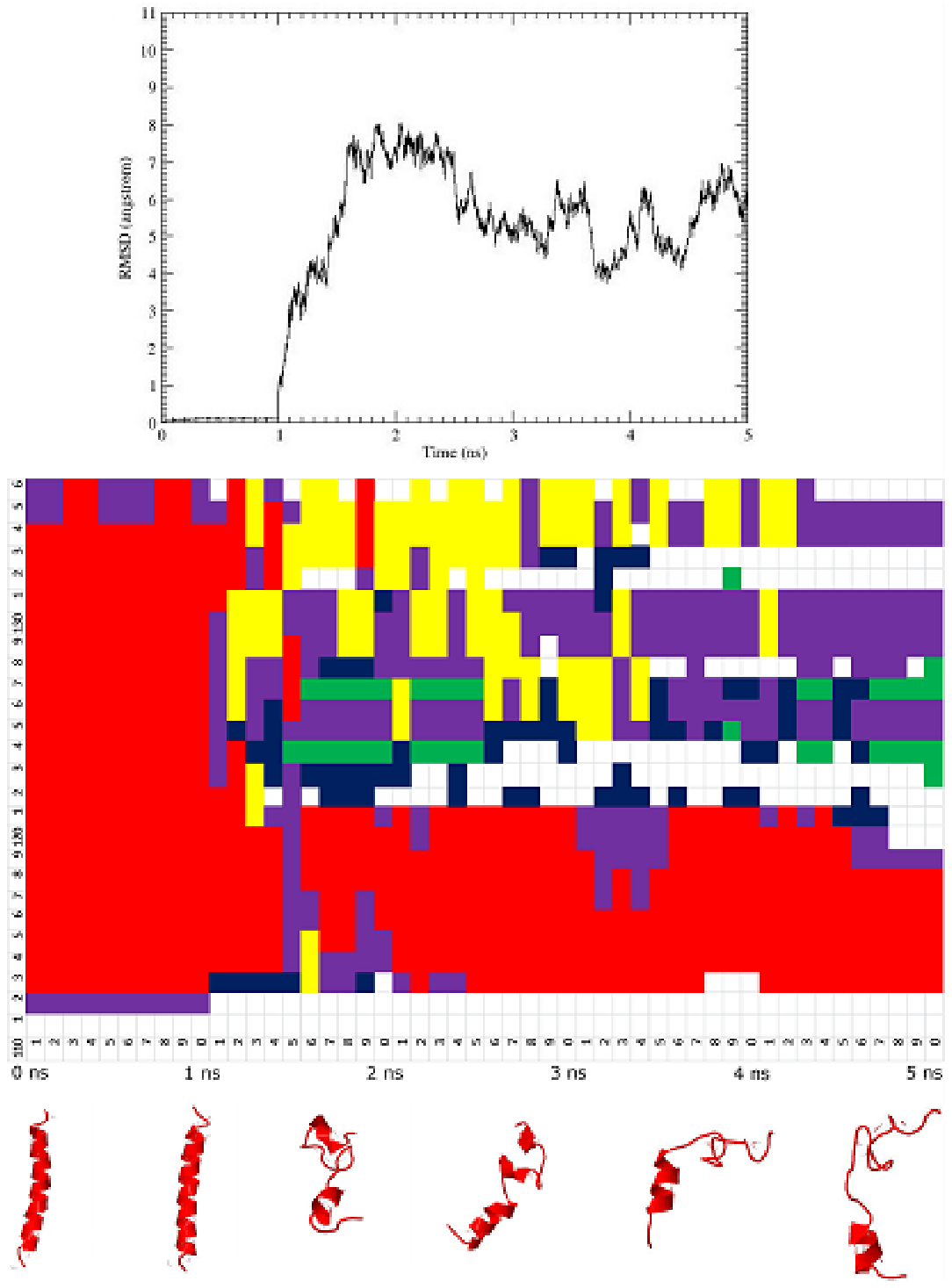}
}
\caption{{\bf Mutant G119L} Variations of (A) the root-mean-squared deviations (RMSD), (B) the secondary structures, and (C) the respective snapshots at 0 ns, 1 ns, 2 ns, 3 ns, 4 ns and 5 ns of the MD simulations for mutant G119L.}
\end{figure}

\begin{figure}[h!] \label{Fig07}
\centerline{
\includegraphics[width=6.4in]{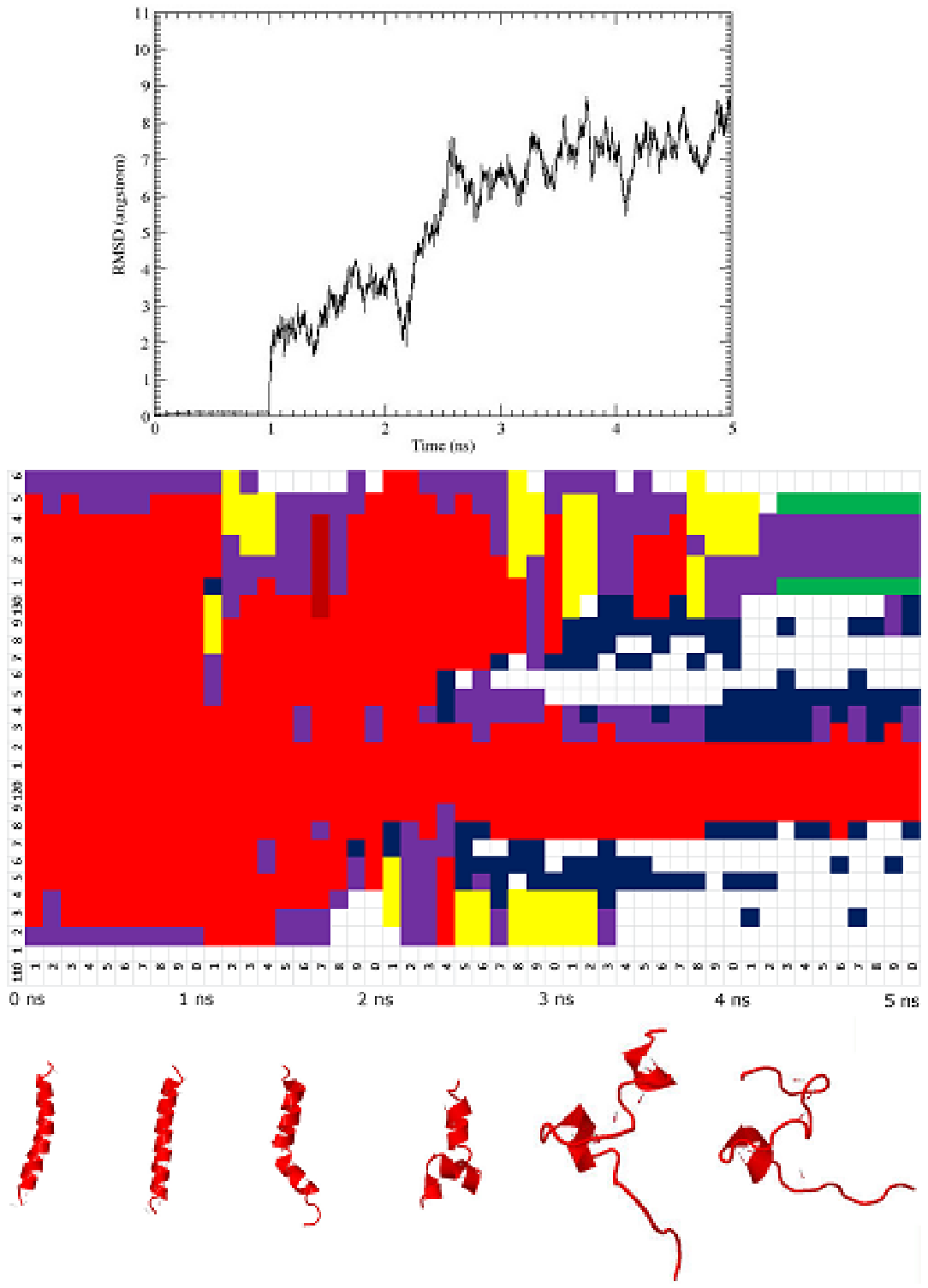}
}
\caption{{\bf Mutant G119P} Variations of (A) the root-mean-squared deviations (RMSD), (B) the secondary structures, and (C) the respective snapshots at 0 ns, 1 ns, 2 ns, 3 ns, 4 ns and 5 ns of the MD simulations for mutant G119P.}
\end{figure}

\begin{figure}[h!] \label{Fig08}
\centerline{
\includegraphics[width=6.4in]{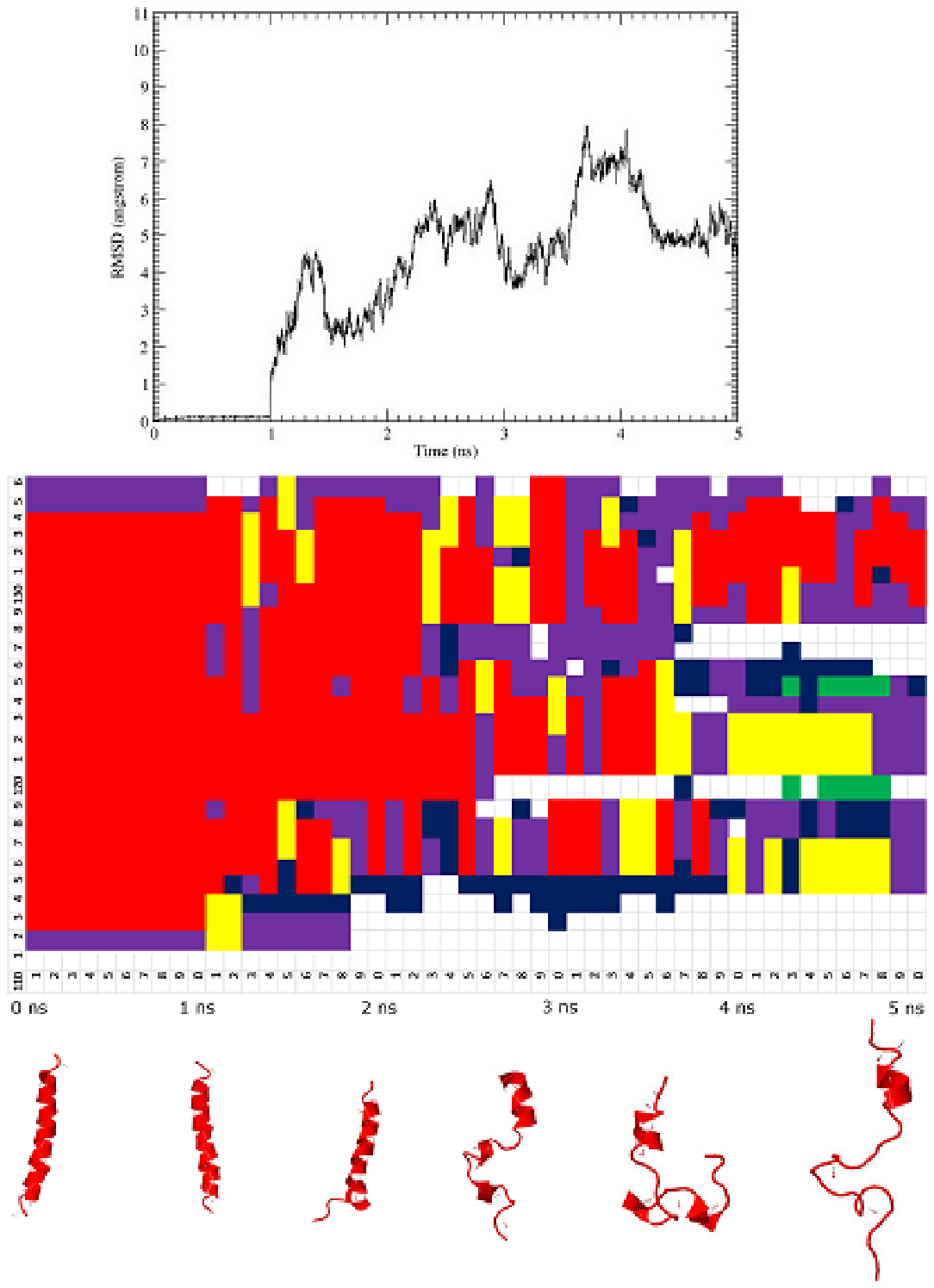}
}
\caption{{\bf Mutant A120P} Variations of (A) the root-mean-squared deviations (RMSD), (B) the secondary structures, and (C) the respective snapshots at 0 ns, 1 ns, 2 ns, 3 ns, 4 ns and 5 ns of the MD simulations for mutant A120P.}
\end{figure}

\begin{figure}[h!] \label{Fig09}
\centerline{
\includegraphics[width=6.4in]{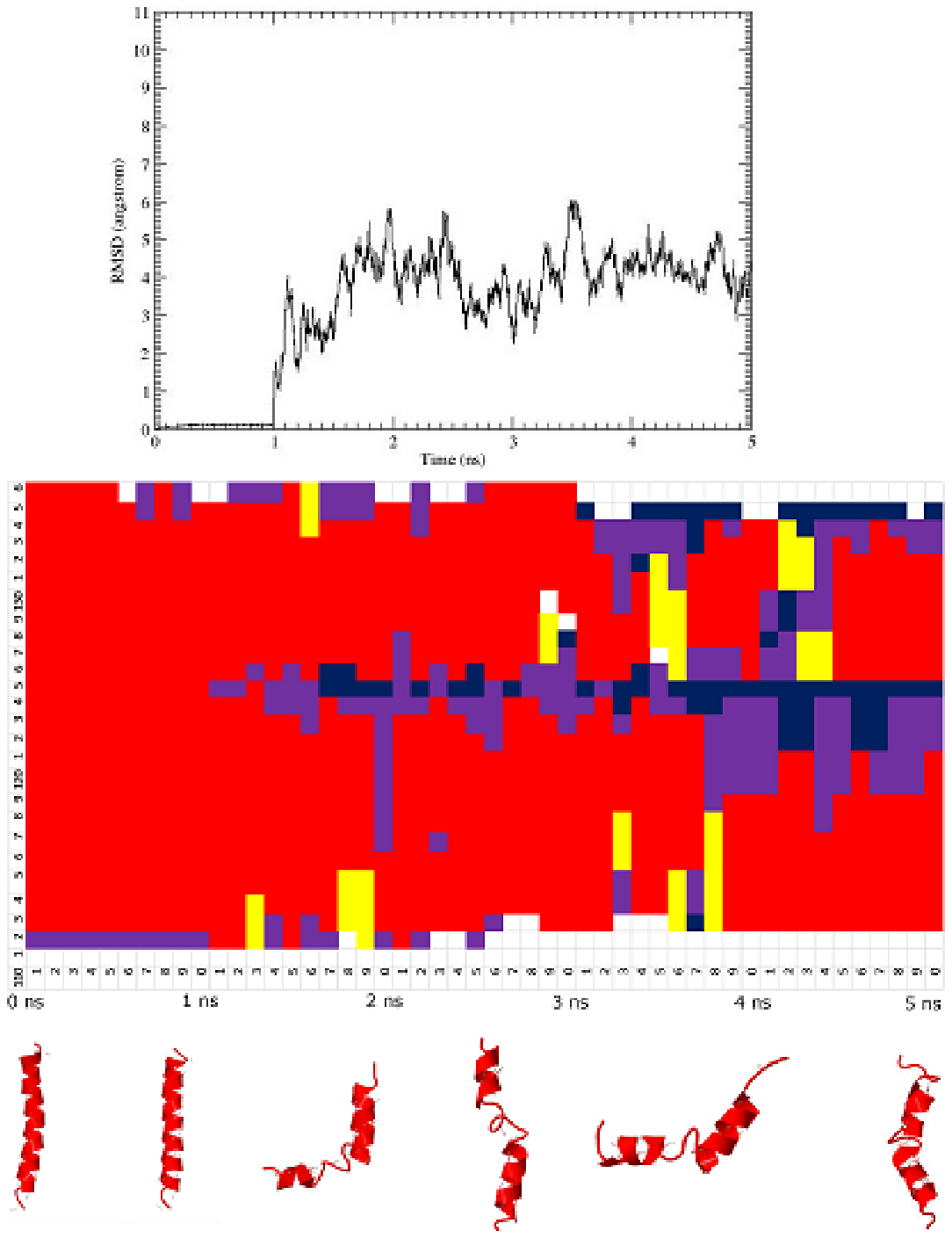}
}
\caption{{\bf Mutant G123A} Variations of (A) the root-mean-squared deviations (RMSD), (B) the secondary structures, and (C) the respective snapshots at 0 ns, 1 ns, 2 ns, 3 ns, 4 ns and 5 ns of the MD simulations for mutant G123A.}
\end{figure}

\begin{figure}[h!] \label{Fig10}
\centerline{
\includegraphics[width=6.4in]{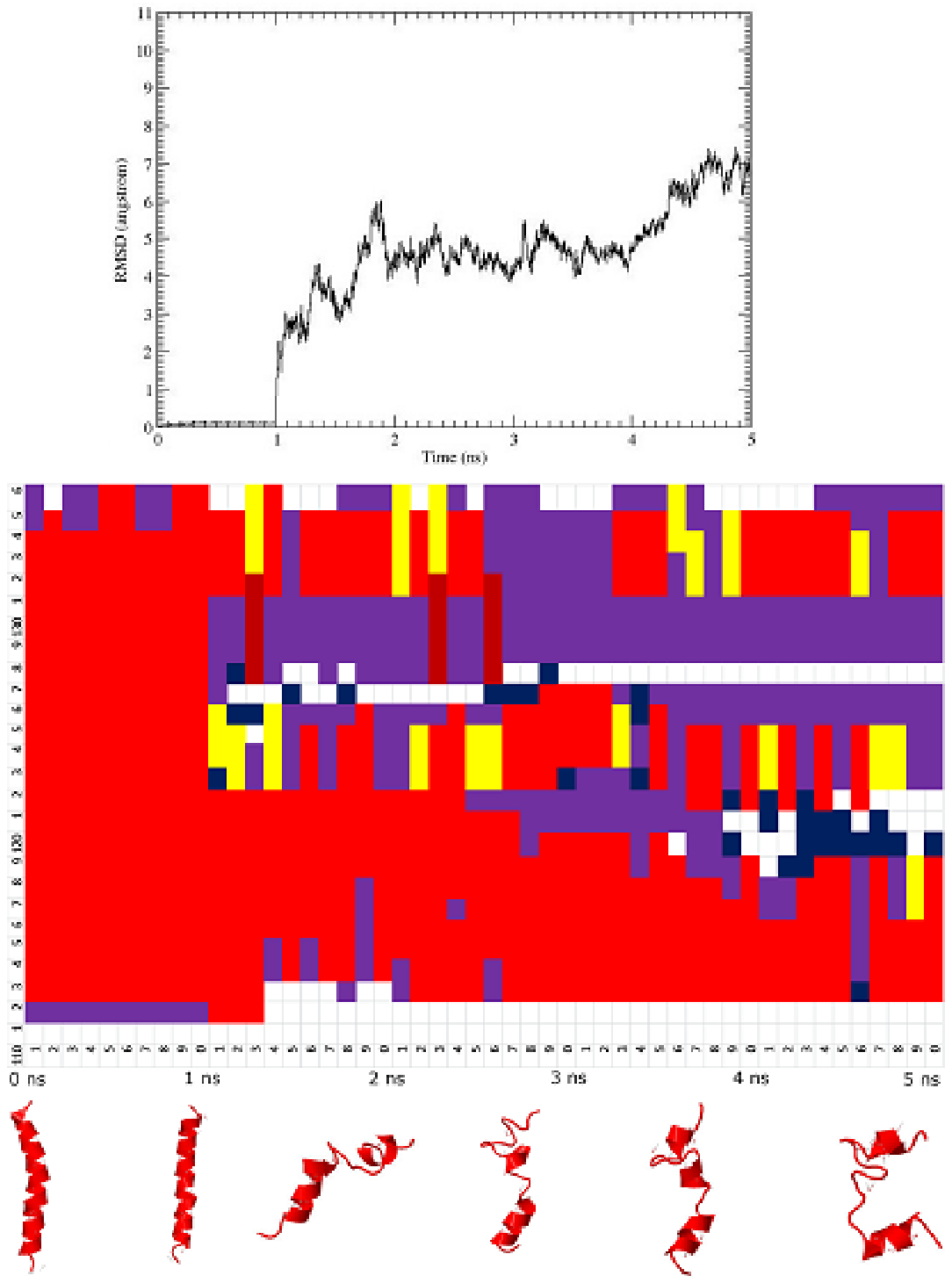}
}
\caption{{\bf Mutant G123P} Variations of (A) the root-mean-squared deviations (RMSD), (B) the secondary structures, and (C) the respective snapshots at 0 ns, 1 ns, 2 ns, 3 ns, 4 ns and 5 ns of the MD simulations for mutant G123P.}
\end{figure}

\begin{figure}[h!] \label{Fig11}
\centerline{
\includegraphics[width=6.4in]{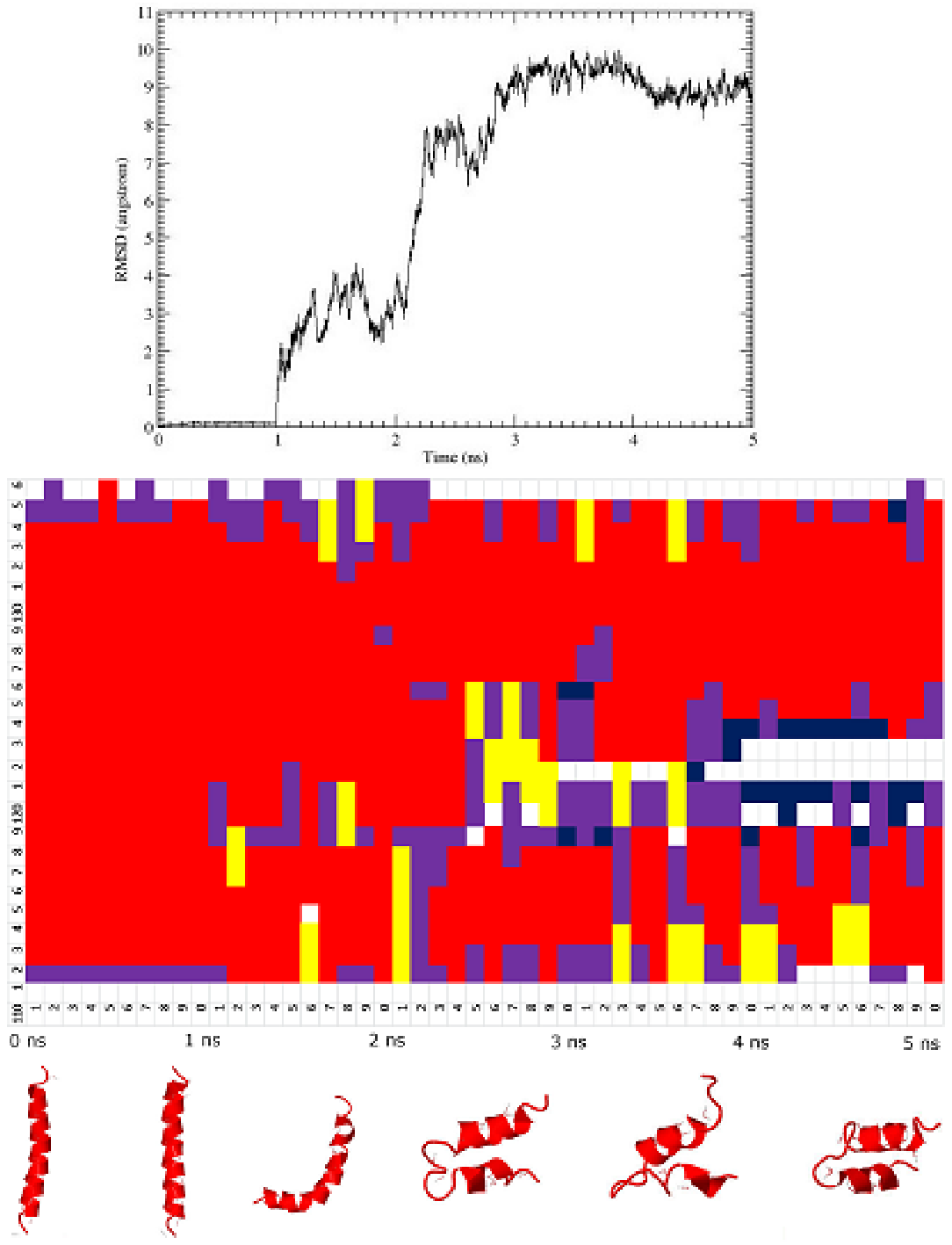}
}
\caption{{\bf Mutant G124A} Variations of (A) the root-mean-squared deviations (RMSD), (B) the secondary structures, and (C) the respective snapshots at 0 ns, 1 ns, 2 ns, 3 ns, 4 ns and 5 ns of the MD simulations for mutant G124A.}
\end{figure}

\begin{figure}[h!] \label{Fig12}
\centerline{
\includegraphics[width=6.4in]{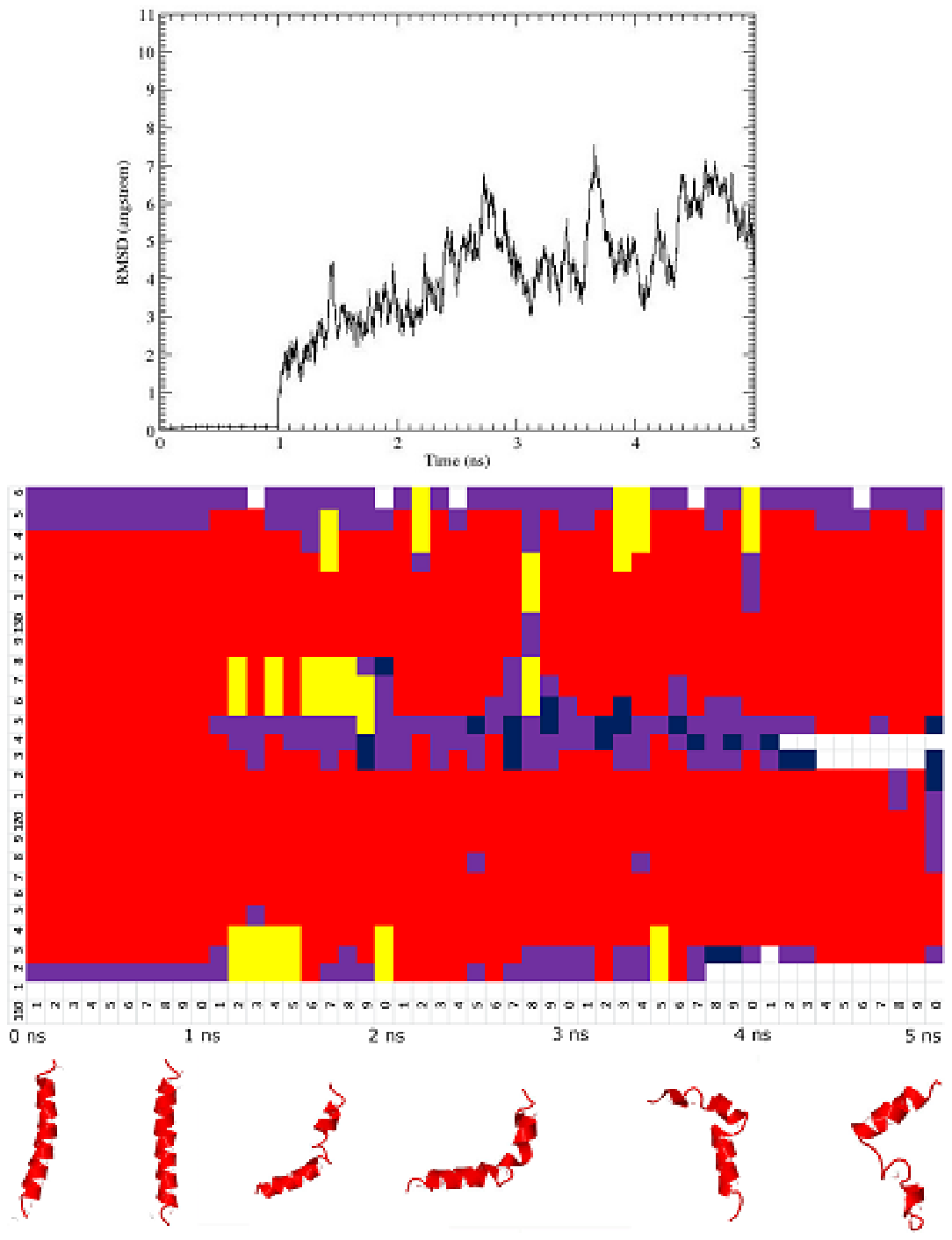}
}
\caption{{\bf Mutant L125A} Variations of (A) the root-mean-squared deviations (RMSD), (B) the secondary structures, and (C) the respective snapshots at 0 ns, 1 ns, 2 ns, 3 ns, 4 ns and 5 ns of the MD simulations for mutant L125A.}
\end{figure}

\begin{figure}[h!] \label{Fig13}
\centerline{
\includegraphics[width=6.4in]{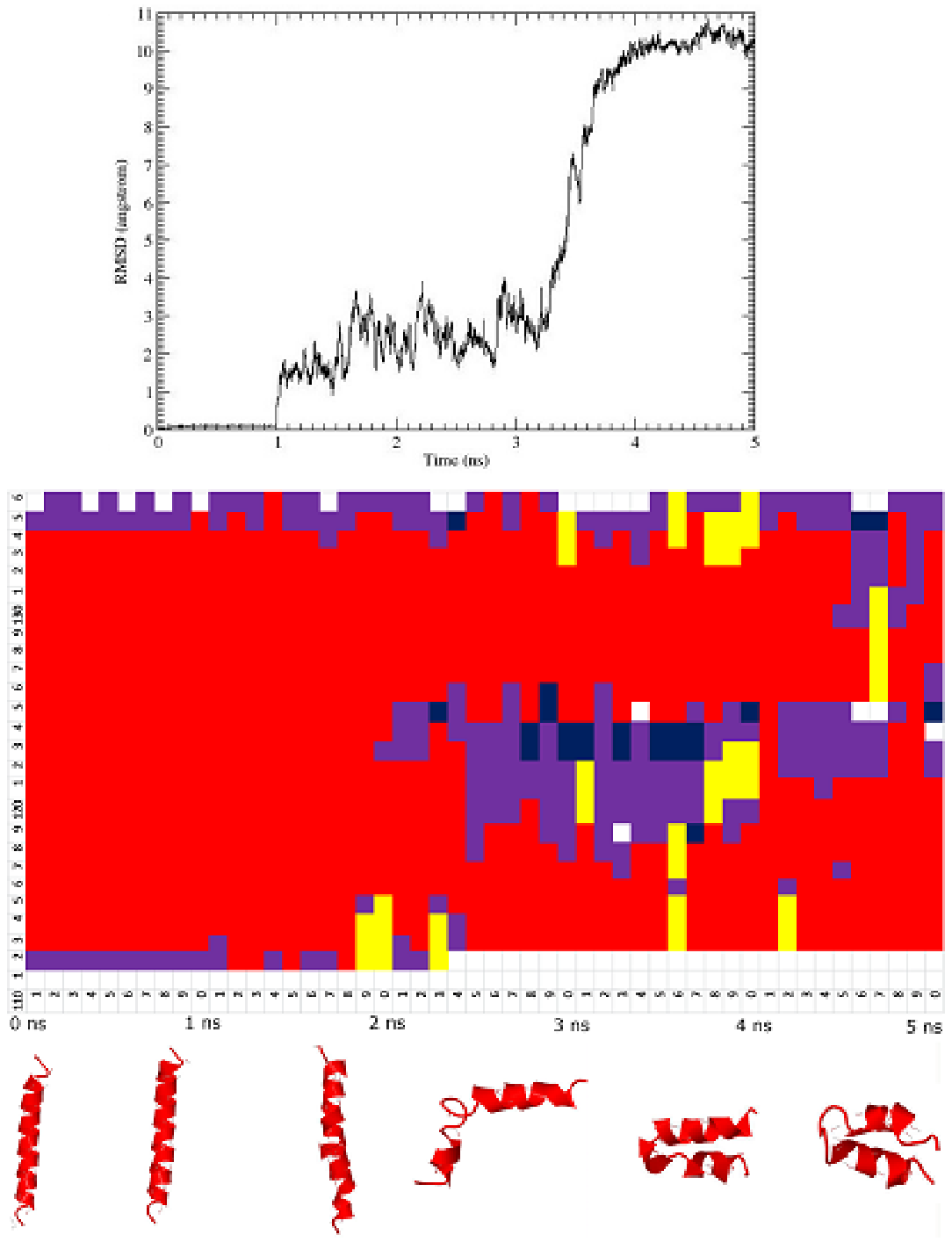}
}
\caption{{\bf Mutant G126A} Variations of (A) the root-mean-squared deviations (RMSD), (B) the secondary structures, and (C) the respective snapshots at 0 ns, 1 ns, 2 ns, 3 ns, 4 ns and 5 ns of the MD simulations for mutant G126A.}
\end{figure}

\begin{figure}[h!] \label{Fig14}
\centerline{
\includegraphics[width=6.4in]{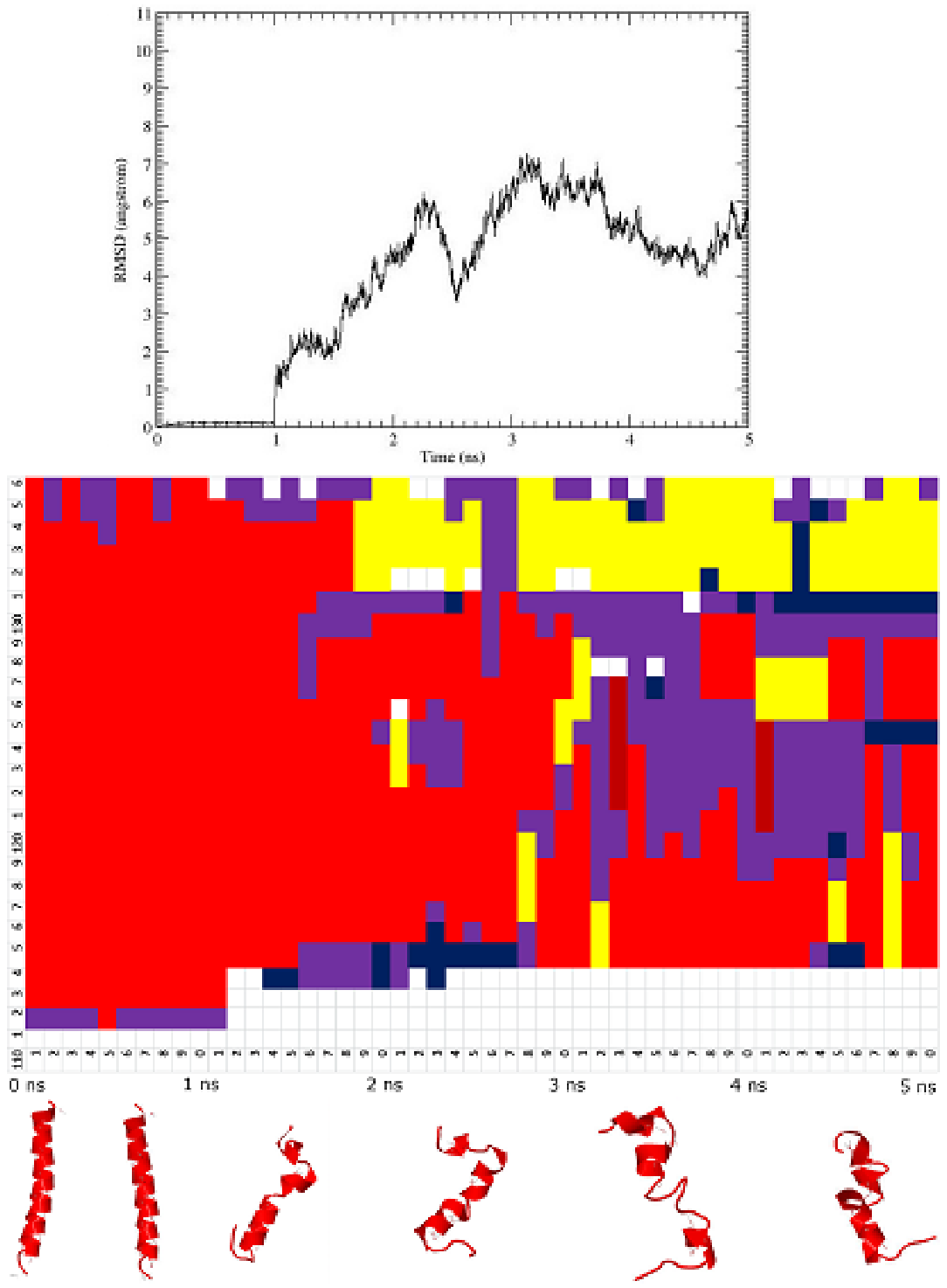}
}
\caption{{\bf Mutant G127A} Variations of (A) the root-mean-squared deviations (RMSD), (B) the secondary structures, and (C) the respective snapshots at 0 ns, 1 ns, 2 ns, 3 ns, 4 ns and 5 ns of the MD simulations for mutant G127A.}
\end{figure}

\begin{figure}[h!] \label{Fig15}
\centerline{
\includegraphics[width=6.4in]{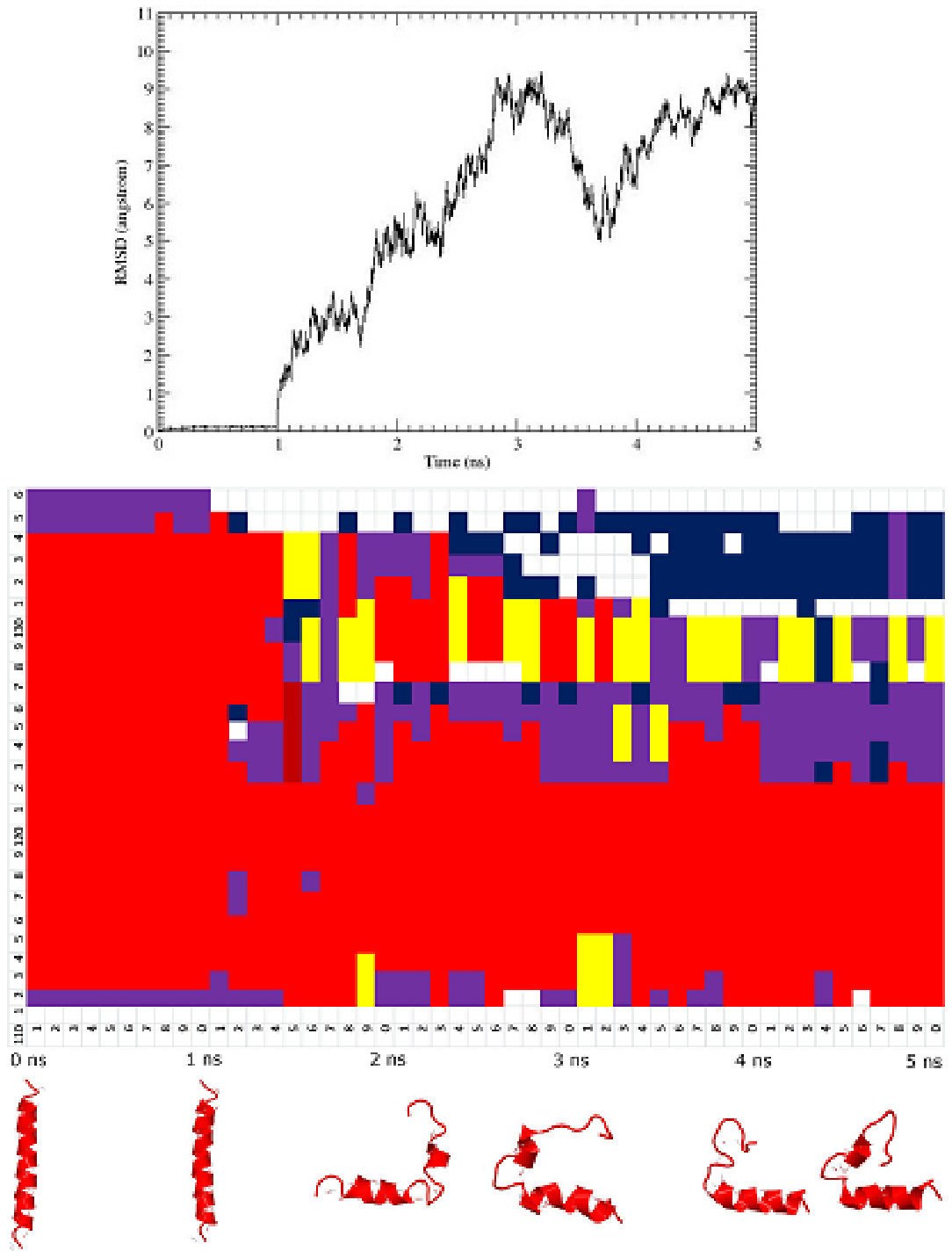}
}
\caption{{\bf Mutant G127L} Variations of (A) the root-mean-squared deviations (RMSD), (B) the secondary structures, and (C) the respective snapshots at 0 ns, 1 ns, 2 ns, 3 ns, 4 ns and 5 ns of the MD simulations for mutant G127L.}
\end{figure}

\begin{figure}[h!] \label{Fig16}
\centerline{
\includegraphics[width=6.4in]{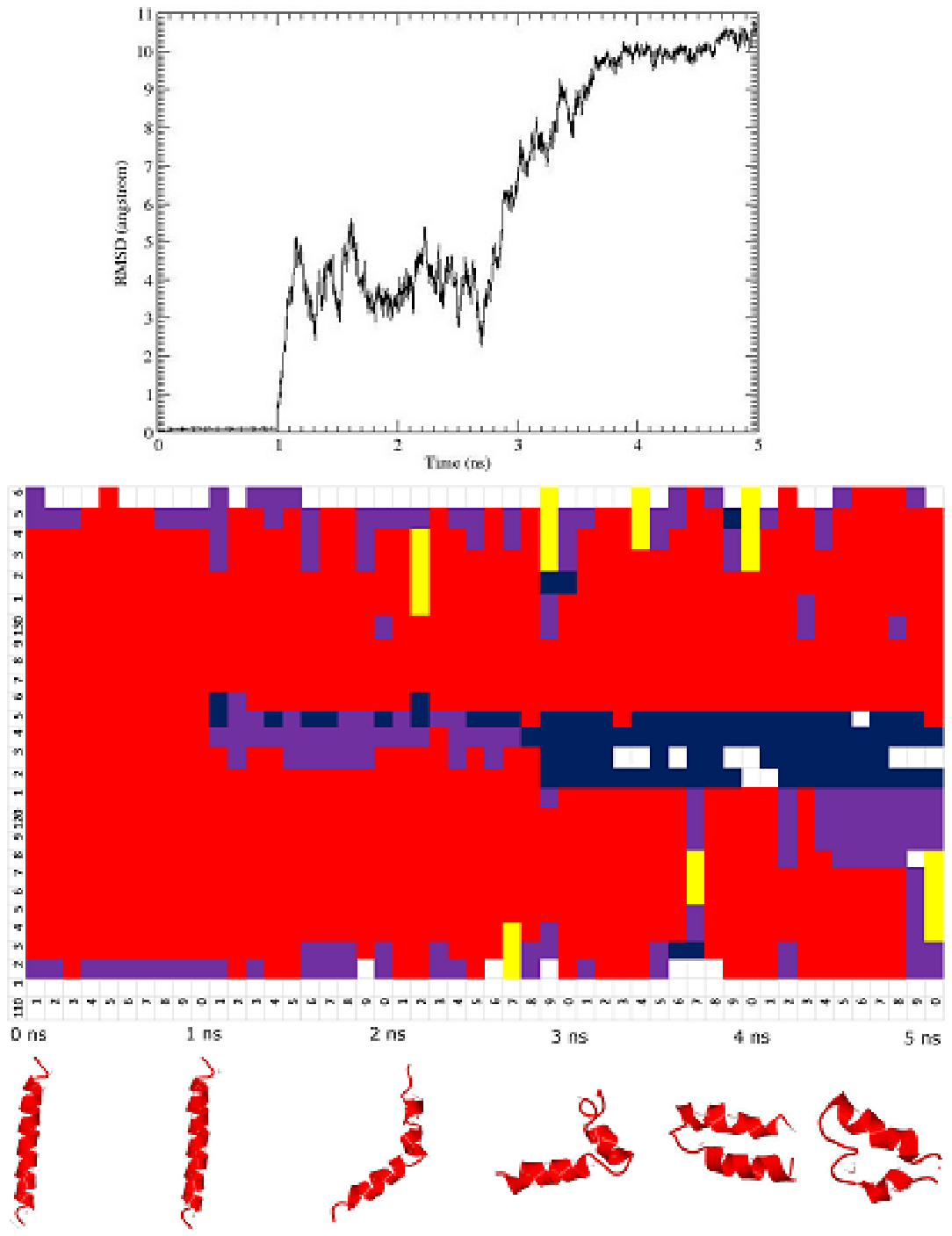}
}
\caption{{\bf Mutant M129V} Variations of (A) the root-mean-squared deviations (RMSD), (B) the secondary structures, and (C) the respective snapshots at 0 ns, 1 ns, 2 ns, 3 ns, 4 ns and 5 ns of the MD simulations for mutant M129V.}
\end{figure}

\begin{figure}[h!] \label{Fig17}
\centerline{
\includegraphics[width=6.4in]{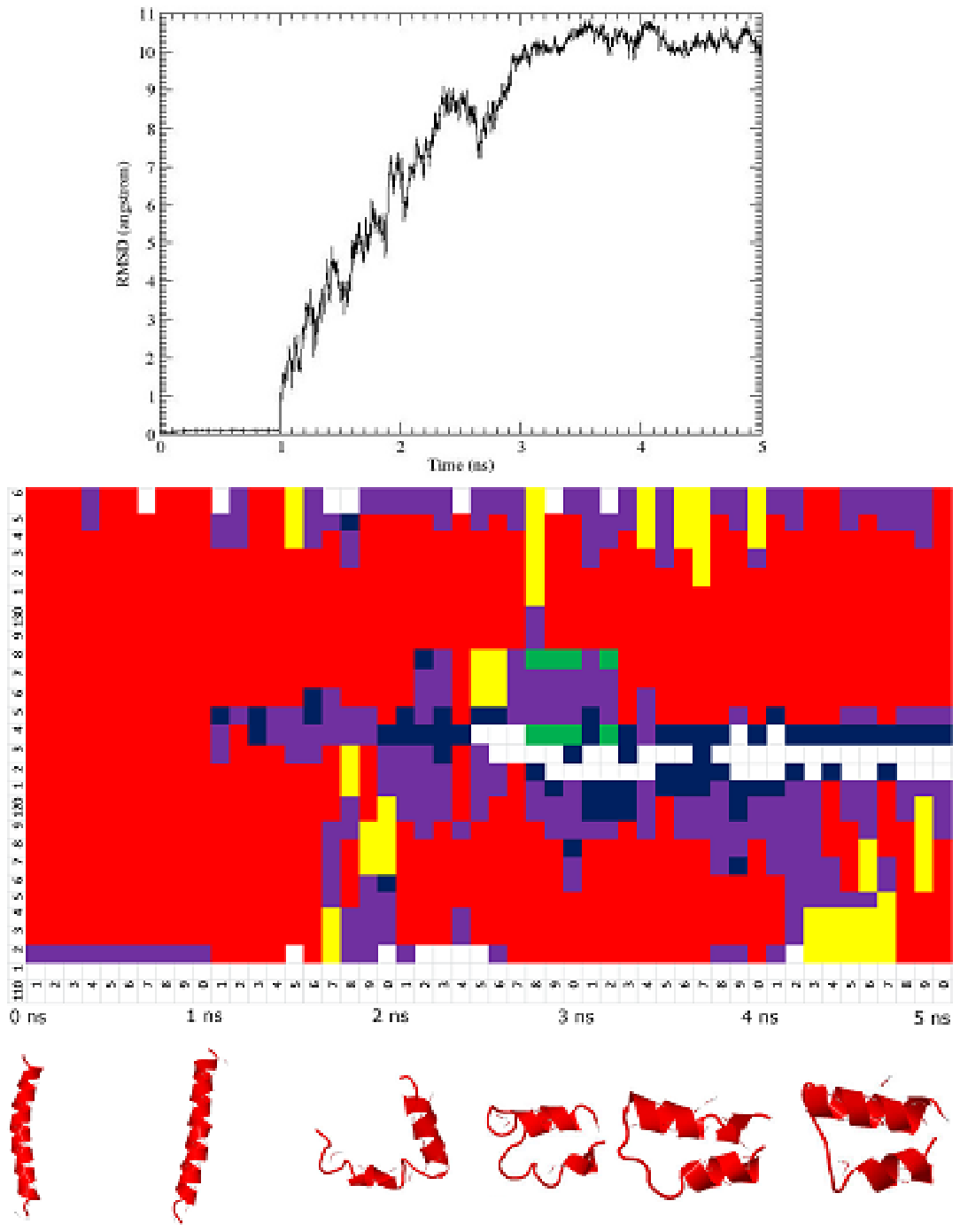}
}
\caption{{\bf Mutant G131A} Variations of (A) the root-mean-squared deviations (RMSD), (B) the secondary structures, and (C) the respective snapshots at 0 ns, 1 ns, 2 ns, 3 ns, 4 ns and 5 ns of the MD simulations for mutant G131A.}
\end{figure}

\begin{figure}[h!] \label{Fig18}
\centerline{
\includegraphics[width=6.4in]{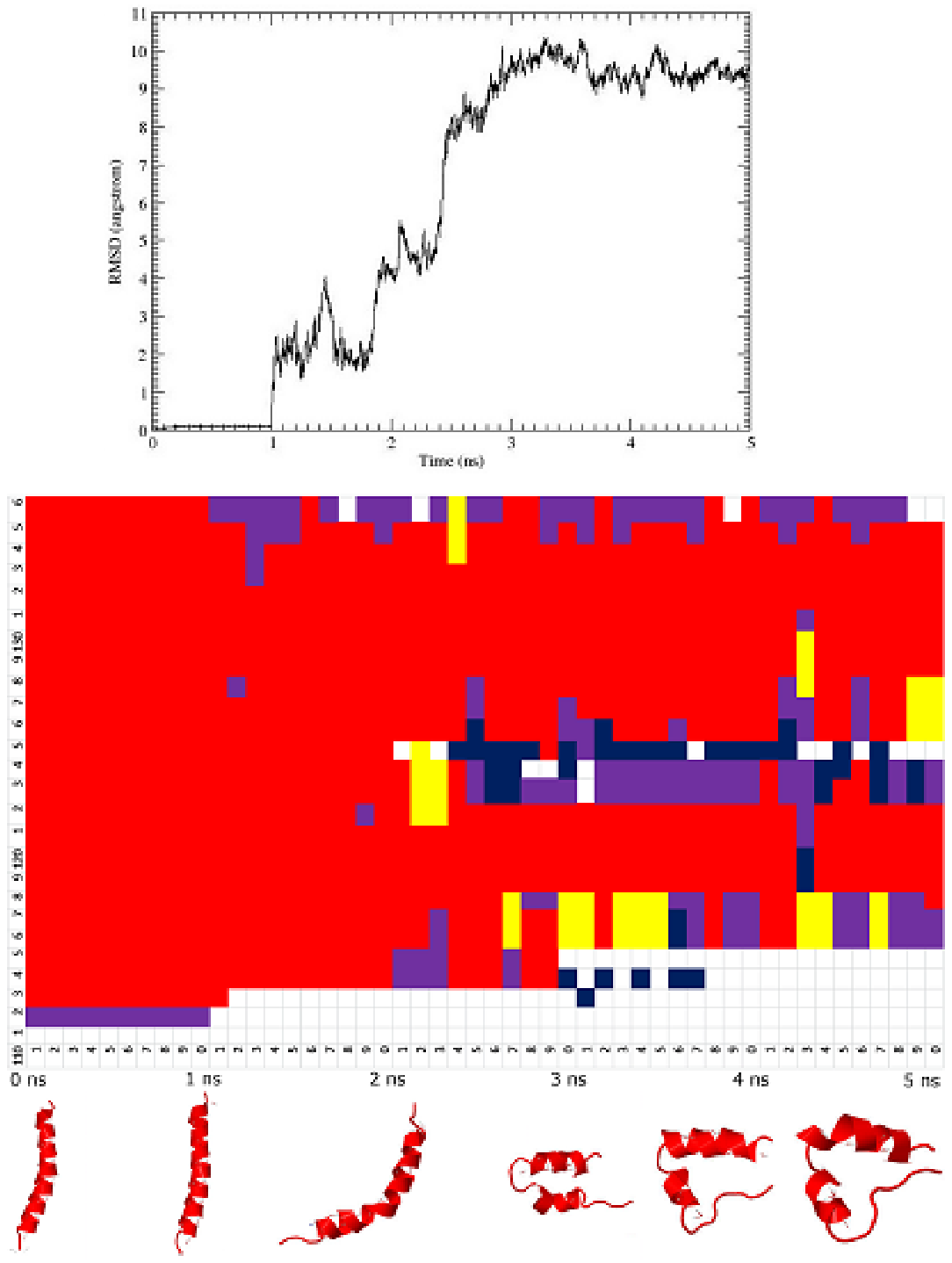}
}
\caption{{\bf Mutant G131L} Variations of (A) the root-mean-squared deviations (RMSD), (B) the secondary structures, and (C) the respective snapshots at 0 ns, 1 ns, 2 ns, 3 ns, 4 ns and 5 ns of the MD simulations for mutant G131L.}
\end{figure}

\begin{figure}[h!] \label{Fig19}
\centerline{
\includegraphics[width=6.4in]{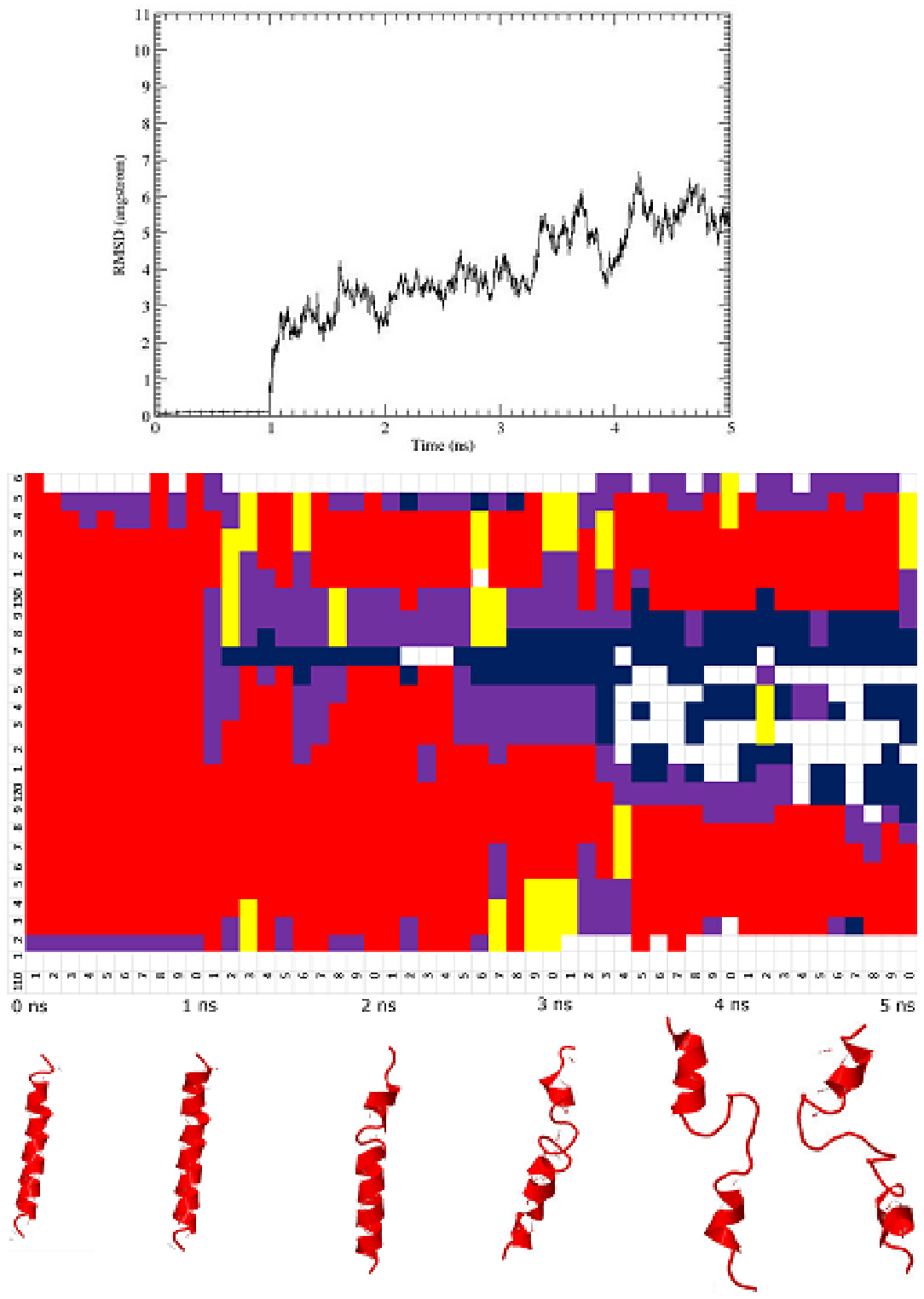}
}
\caption{{\bf Mutant G131P} Variations of (A) the root-mean-squared deviations (RMSD), (B) the secondary structures, and (C) the respective snapshots at 0 ns, 1 ns, 2 ns, 3 ns, 4 ns and 5 ns of the MD simulations for mutant G131P.}
\end{figure}
\end{document}